%% file: arXiv.tex
\documentclass[peerreview]{IEEEtran}

\usepackage{hyperref} %
\hypersetup{
    colorlinks,
    linkcolor={blue!80!black},%
    citecolor={green!30!black},
    urlcolor={blue!80!black}
}

\usepackage[bottom]{footmisc}

\usepackage{fontawesome5}
\usepackage{tikzpeople}

\usepackage{enumerate}
\usepackage{amsmath,amssymb,amsfonts}
\usepackage{bbm}
\usepackage{nicefrac}
\usepackage{amsthm}
\usepackage{algorithmic}
\usepackage{cleveref}
\usepackage{textcomp}
\def\BibTeX{{\rm B\kern-.05em{\sc i\kern-.025em b}\kern-.08em
T\kern-.1667em\lower.7ex\hbox{E}\kern-.125emX}}
\usepackage{physics}
\usepackage{multirow}
\usepackage{siunitx}
\AtBeginDocument{\RenewCommandCopy\qty\SI}
\ExplSyntaxOn
\msg_redirect_name:nnn { siunitx } { physics-pkg } { none }
\ExplSyntaxOff
\usepackage{float}
\usepackage{graphicx}
\usepackage{xcolor}
\usepackage{comment}
\renewcommand{\trace}{\mathrm{Tr}}
\AtBeginDocument{\RenewCommandCopy\qty\SI}
\theoremstyle{remark}	\newtheorem{theorem}{Theorem}
\theoremstyle{remark}	
\theoremstyle{remark}	\newtheorem{corollary}[theorem]{Corollary}
\theoremstyle{remark}	
\theoremstyle{remark} 
\theoremstyle{remark} \newtheorem{remark}{Remark}
\theoremstyle{remark} \newtheorem{example}{Example}

\newcommand{\identity}{\mathbbm{1}}

\renewcommand{\trace}{\mathrm{Tr}}

\usepackage{tikz}
\usetikzlibrary{shapes,arrows,arrows.meta}
\tikzstyle{line}=[draw] 
\usepackage{tikzpagenodes}
\usepackage{qcircuit}
\usetikzlibrary{math}
\usepackage[square,numbers,sort&compress]{natbib}

\usetikzlibrary{positioning,decorations.pathmorphing}

\input{tikz_pack.tex}

\allowdisplaybreaks 

\title{Quantum Coordination and Nonlocal Games: Theory and Applications}

\author{Husein Natur and Uzi Pereg}
\begin{document}
\maketitle
\begin{abstract}
Coordination is a fundamental primitive in communication and information theory, in which distributed systems must collectively generate correlated behavior rather than merely exchange messages. In quantum networks, the nature of entanglement, quantum measurements, and nonclassical correlations introduces coordination possibilities unavailable classically. This article reviews recent advances in coordination over classical and quantum communication networks, focusing on empirical and strong coordination in multi-user settings.

We consider the implications of coordination for nonlocal games, showing how it provides a natural framework for understanding the correlations that enable spatially separated players to improve their probability of winning. 
We present a unified framework for coordination using classical or quantum communication and pre-shared correlation resources.
 The review covers simulation of both entanglement and separable 
correlations across a variety of network architectures, including two-node, cascade, broadcast, and multiple-access networks. We study the operational differences between empirical and strong coordination, and the tradeoffs between communication and correlation resources, such as pre-shared randomness and entanglement.

Coordination plays a major role in device-independent quantum key distribution (DI-QKD) schemes, in which parties can generate a secret key even if the devices used in the process have been prepared by an adversary. 
Furthermore, we examine the role of coordination in quantum repeaters, where distributed entanglement serves as a resource for long-distance quantum communication.
The review  highlights connections between coordination theory and applications such as
distributed quantum systems, quantum internet architectures, and future communication networks. 
\end{abstract}

\section{Introduction}
\label{Section: Introduction}

Modern communication networks increasingly support distributed tasks in which several users must act jointly rather than merely exchange messages between isolated source-destination pairs \cite{cuff2010coordination,sudan2019communication}. Coordination and cooperation arise in applications such as distributed sensing \cite{stankovic2003real}, autonomous vehicles \cite{ahangar2021survey}, cooperative games \cite{cuff2011coordination}, distributed computation \cite{borcea2002cooperative}, machine-to-machine communication \cite{mylonakis2020remote}, the Internet of Things \cite{he2020internet,torres2023message}, and secure communication \cite{10619085}. These settings involve multiple terminals operating over shared and limited resources, and their performance often depends on the collective behavior of the network. Network information theory provides a general framework for studying such multi-terminal communication systems \cite{ElGamalKim:11b,tse1999linear,rosenberger2023identification,pereg2023multiple,8768393}, while coordination theory provides a natural perspective for studying their joint operation \cite{le2017joint}.

In classical networks, coordination concerns the generation of a prescribed joint behavior among spatially separated users. The terminals may observe correlated sources, exchange messages, and use shared randomness in order to produce actions whose joint distribution approximates a desired target distribution \cite{cuff2010coordination,sudan2019communication,le2017joint}. The main objective is to characterize the communication and randomness resources required to establish this correlation. This formulation applies to networks in which the users must operate consistently, make related decisions, or generate coordinated outputs without necessarily reconstructing the information observed by the other terminals.
The coordination framework becomes richer in quantum networks, where the terminals may additionally exchange quantum systems and share entanglement. The target is no longer restricted to a classical joint distribution and may instead consist of a joint quantum state, correlated measurement outcomes, or a prescribed input-output behavior. Quantum resources can generate correlations that have no classical counterpart and are expected to play an important role in future quantum networks \cite{notzel2020entanglement,9232550,burenkov2021practical,granelli2022novel}. Quantum coordination therefore studies the communication, randomness, and entanglement resources required to generate a desired quantum state or observable behavior among spatially separated users.

A simple cascade network already illustrates an important difference between classical and quantum coordination. Consider a network $A\to B\to C$, where Alice communicates with Bob, who then communicates with Charlie. For the classical cascade model considered here, the communication rate from Alice to Bob is at least as large as the rate from Bob to Charlie, since the information delivered to Charlie must first reach Bob.

In the quantum setting, this ordering may be reversed. When the target state is entangled, Alice may send fewer qubits to Bob than Bob sends to Charlie \cite{natur2025quantum}. This is possible because a globally pure entangled state can have mixed reduced states. Consequently, the communication required on each link depends not only on the local subsystem being generated, but also on its correlations with the rest of the network. We will visit this example again in detail in Subsection~\ref{Subsubsection: Cascade Network}. This behavior is closely related to the fact that quantum conditional entropy can be negative.
More generally, several properties of quantum mechanics distinguish coordination with quantum communication from coordination with classical communication.

\subsection{Applications}
We list several applications that require coordination protocols.
\label{Subsection: Applications}
\subsubsection{Sensing networks}
\begin{figure}[t]
\centering

\begin{tikzpicture}[
    peerlink/.style={
        <->,
        >=Latex,
        draw=blue!60,
        line width=0.8pt,
        dashed,
        shorten <=5pt,
        shorten >=5pt
    },
    sensing/.style={
        draw=cyan!35,
        fill=cyan!5,
        dashed,
        line width=0.6pt
    },
    agent/.style={
        inner sep=0pt
    }
]

\coordinate (C1pos) at (-3.4,0.5);
\coordinate (C2pos) at (0,-1.0);
\coordinate (C3pos) at (3.4,0.5);
\coordinate (Dpos)  at (0.3,3.7);

\draw[peerlink]
    (C1pos) to[bend right=10] (C2pos);

\draw[peerlink]
    (C2pos) to[bend right=10] (C3pos);

\draw[peerlink]
    (C1pos) to[bend left=13] (C3pos);

\draw[peerlink]
    (Dpos) to[bend right=8] (C1pos);

\draw[peerlink]
    (Dpos) to[bend right=8] (C2pos);

\draw[peerlink]
    (Dpos) to[bend left=8] (C3pos);

\node[agent] (Car1) at (C1pos)
{
    \includegraphics[
        width=2.15cm,
        angle=0,
        origin=c
    ]{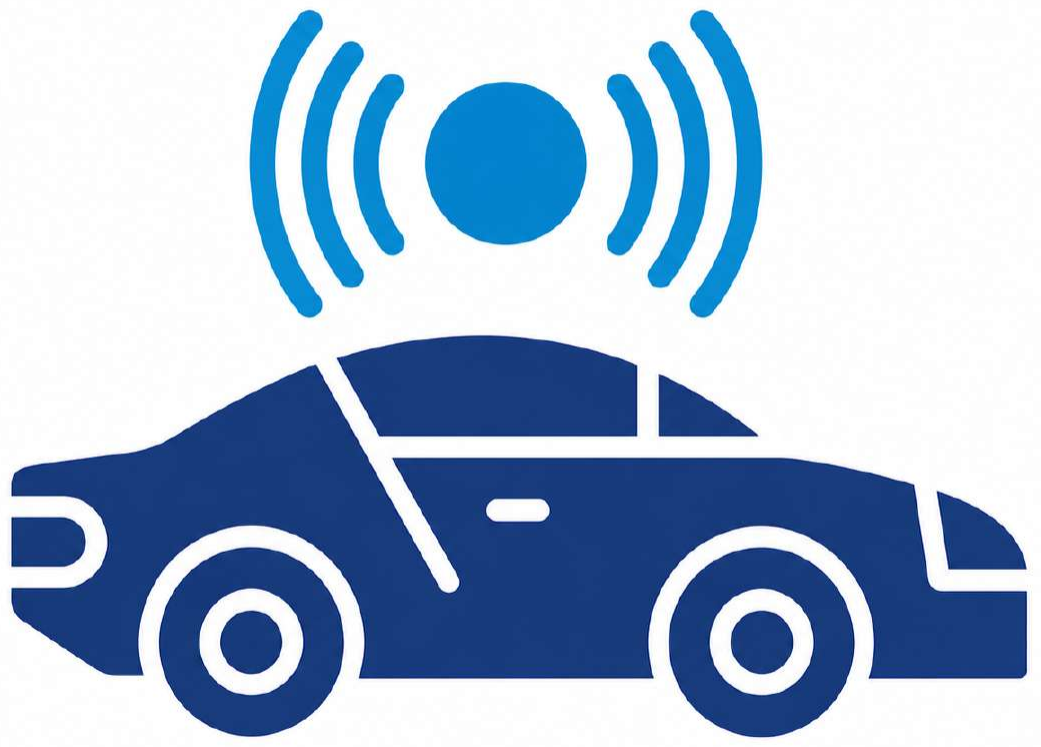}
};

\node[agent] (Car2) at (C2pos)
{
    \includegraphics[
        width=2.15cm
    ]{Car.pdf}
};

\node[agent] (Car3) at (C3pos)
{
    \includegraphics[
        width=2.15cm,
        angle=0,
        origin=c
    ]{Car.pdf}
};

\node[agent] (Drone) at (Dpos)
{
    \includegraphics[
        width=1.75cm,
        angle=0,
        origin=c
    ]{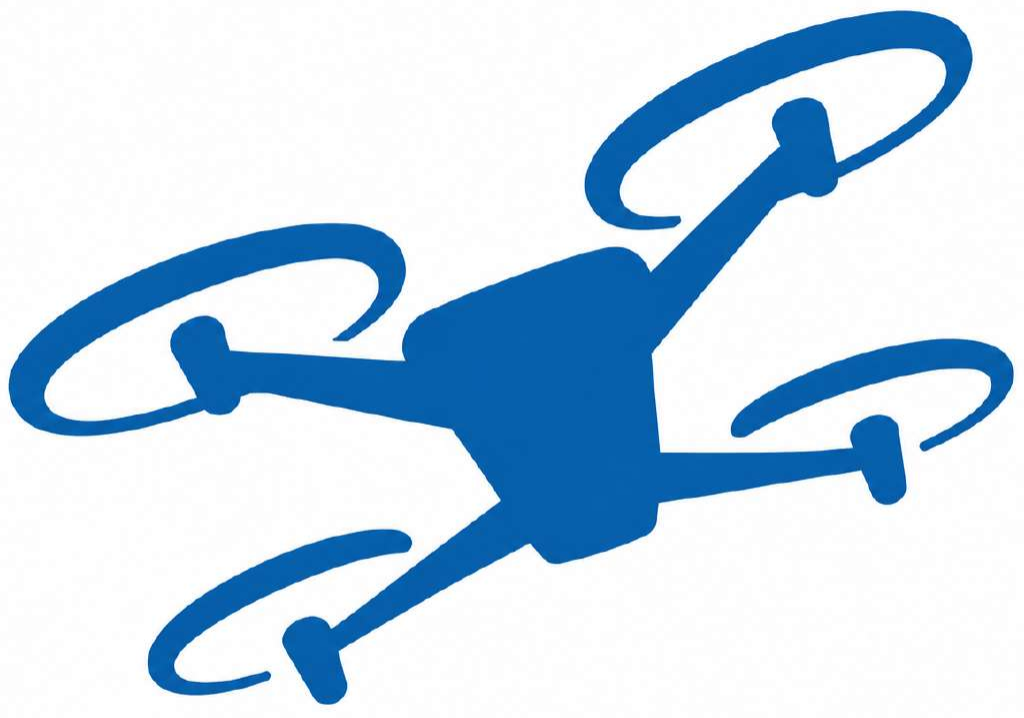}
};

\end{tikzpicture}

\caption{
Coordination in an autonomous network.
Each autonomous vehicle and drone observes only a local part of the
environment and exchanges limited information directly with neighboring
agents. The communication links allow the agents to coordinate
their actions without relying on a central control unit.
}
\label{Figure: Sensing}

\end{figure}

In distributed sensing networks, 
multiple sensors or agents observe different parts of an environment and exchange limited information in order to collectively estimate a parameter, detect an event, or make a joint decision. 
The objective is therefore not merely to communicate local observations reliably, but rather to generate a coordinated global behavior from distributed information. 
This paradigm also appears in cooperative control and autonomous systems, where agents must synchronize actions and adapt to changing environments using only partial local information, see Figure~\ref{Figure: Sensing}. 
Quantum networks introduce additional coordination resources through entanglement and quantum communication links, enabling distributed users to establish correlations that are impossible to reproduce classically \cite{HsiehWilde:10p1}.
For example, entangled quantum sensors can improve the precision of distributed measurements \cite{shapiro2020quantum}, while quantum coordination protocols generate multipartite correlations required for distributed quantum tasks.

\subsubsection{Games}
\begin{figure*}[t]
\centering

\begin{tikzpicture}[
    player/.style={
        inner sep=2pt,
        fill=white
    },
    coordinationLink/.style={
        draw=black!65!white,
        line width=0.9pt,
        shorten <=6pt,
        shorten >=6pt
    },
    refereeLink/.style={
        <->,
        >=Latex,
        draw=black!65!white,
        line width=0.9pt,
        shorten <=6pt,
        shorten >=6pt
    },
    transition/.style={
        ->,
        >=Latex,
        draw=orange!85!black,
        line width=1.2pt
    }
]

\begin{scope}[xshift=-4.5cm]

\coordinate (LP1) at (0,2.2);
\coordinate (LP2) at (2.3,0);
\coordinate (LP3) at (0,-2.2);
\coordinate (LP4) at (-2.3,0);

\draw[coordinationLink]
    (LP1) to[bend left=18] (LP2);

\draw[coordinationLink]
    (LP2) to[bend left=18] (LP3);

\draw[coordinationLink]
    (LP3) to[bend left=18] (LP4);

\draw[coordinationLink]
    (LP4) to[bend left=18] (LP1);

\draw[coordinationLink]
    (LP1) -- (LP3);

\draw[coordinationLink]
    (LP4) -- (LP2);

\node[player] at (LP1)
{
    \includegraphics[
        height=1.25cm
    ]{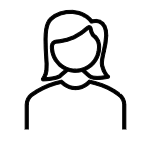}
};

\node[player] at (LP2)
{
    \includegraphics[
        height=1.25cm
    ]{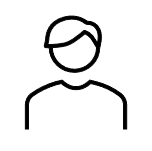}
};

\node[player] at (LP3)
{
    \includegraphics[
        height=1.4cm
    ]{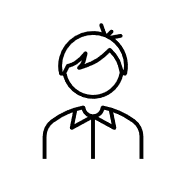}
};

\node[player] at (LP4)
{
    \includegraphics[
        height=1.4cm
    ]{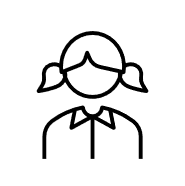}
};

\node[
    font=\small,
    align=center
]
at (0,-3)
{(a) Coordination before the game};

\end{scope}

\draw[transition]
    (-0.9,0) -- (0.9,0);

\begin{scope}[xshift=4.5cm]

\coordinate (RP1) at (0,2.2);
\coordinate (RP2) at (2.3,0);
\coordinate (RP3) at (0,-2.2);
\coordinate (RP4) at (-2.3,0);

\coordinate (R) at (0,0);

\draw[refereeLink] (R) -- (RP1);
\draw[refereeLink] (R) -- (RP2);
\draw[refereeLink] (R) -- (RP3);
\draw[refereeLink] (R) -- (RP4);


\node[player] at (R)
{
    \includegraphics[
        height=1.25cm
    ]{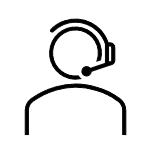}
};

\node[player] at (RP1)
{
    \includegraphics[
        height=1.25cm
    ]{Player_woman.png}
};

\node[player] at (RP2)
{
    \includegraphics[
        height=1.25cm
    ]{Player_man.png}
};

\node[player] at (RP3)
{
    \includegraphics[
        height=1.4cm
    ]{Player_boy.png}
};

\node[player] at (RP4)
{
    \includegraphics[
        height=1.4cm
    ]{Player_girl.png}
};

\node[
    font=\small,
    align=center
]
at (0,-3)
{(b) Execution of the game};

\end{scope}

\end{tikzpicture}

\caption{
Coordination as a preprocessing step in a multiplayer game.
Before the game begins, the players agree on a common strategy and
establish the classical or quantum correlations required to coordinate
their actions.
During the game, each player receives an individual query from the
referee and returns a response according to the previously established
strategy, without communicating with the other players.
}
\label{Figure: Games}

\end{figure*}

Game theory provides a useful framework for studying coordination among distributed decision makers. 
In non-cooperative games, each player pursues an individual objective, whereas in cooperative games the players share a common goal and seek to coordinate their actions in order to improve their collective performance \cite{myerson1997game}, see Figure~\ref{Figure: Games}. 
The effectiveness of a cooperative strategy depends 
on the correlations that can be established among the players. 
From the perspective of coordination theory, communication and shared resources are used to generate these correlations while respecting the constraints imposed by the underlying network. 
This viewpoint has found applications in distributed control, 
multi-agent systems,  
networked decision making, 
and strategic communication \cite{9706458}. 
Quantum information introduces new forms of correlation through entanglement, giving rise to cooperative strategies that outperform their classical counterparts. 
These advantages are most prominently illustrated in quantum nonlocal games, where the available correlations directly determine the achievable performance of the players \cite{brunner2014bell,buhrman2010nonlocality}.

In a nonlocal game, a referee sends individual queries to two or more spatially separated players. Each player then produces a response. 
The players may cooperate and agree on a (possibly random) strategy before the game begins, and may additionally share correlation resources a priori, such as shared randomness or quantum entanglement. Nonetheless, there is no communication during the execution of the game.
The game is won if the responses satisfy a particular relation with respect to the queries.
In general, the winning probability of the game may depend not only on the strategy that the players agree on, but also on the type of correlation pre-shared between them before the game begins. 
The CHSH game is a canonical example \cite{clauser1969proposed}, where 
 Alice and Bob receive binary questions $a,b\in\{0,1\}$ and return  binary answers $x,y\in\{0,1\}$, respectively. 
They win the game if $x\oplus y=a\land b$, i.e.,
 the parity of their responses is the same as the conjunction of the queries. 
Classically,  the maximum winning probability is $75\%$. This bound holds even if Alice and Bob are provided with unlimited pre-shared randomness. 
However, if the players share an entangled quantum state and perform local measurements, they can achieve a higher winning probability,
of $\approx 85.33\%$.
From the coordination viewpoint, entanglement enables the users to generate non-classical behavior that improves the performance of distributed tasks. 
Therefore, quantum coordination plays an essential role in quantum nonlocal games, serving as a preparation step in which players generate the desired correlations before the game begins, resulting in a higher winning probability. %

Bell inequalities provide a mathematical characterization of the behavior that can be generated using only local strategies and classical correlation, i.e., a pre-shared string of random bits. 
Any behavior 
that admits a classical description must satisfy these inequalities. 
A Bell violation occurs when the measurement statistics exceed the maximum value allowed by a Bell inequality, demonstrating that any local (classical) model cannot reproduce the correlations \cite{bell1964einstein}. 
Nonlocal games provide an operational interpretation of Bell inequalities: the winning probability of a game is directly related to the correlations available to the players. 
For example, in the CHSH game, classical strategies achieve a maximum winning probability of $75\%$, corresponding to the classical Bell bound. Quantum correlations that violate this bound yield a higher winning probability and therefore demonstrate a quantum advantage.
Consequently, Bell violations may be viewed as evidence of non-classical coordination resources that enhance the performance of distributed tasks.
The profound difference between classical and quantum correlations is reflected through Bell violations, which show that certain quantum correlations cannot be explained by a local theory. 


\subsubsection{Cryptography} 
\begin{figure}[t]
\centering

\begin{tikzpicture}[
    agent/.style={
        inner sep=2pt
    },
    communication/.style={
        ->,
        >=Latex,
        draw=black,
        line width=1.0pt,
        shorten <=7pt,
        shorten >=7pt
    },
    eavesdrop/.style={
        ->,
        >=Latex,
        draw=orange!85!black,
        dashed,
        line width=0.9pt,
        shorten >=3pt
    }
]

\coordinate (AlicePos) at (-3,0);
\coordinate (BobPos)   at (3,0);
\coordinate (EvePos)   at (0,-2.3);

\coordinate (Channel) at ($(AlicePos)!0.5!(BobPos)$);

\draw[communication]
    ($(AlicePos)+(0.4,0)$)
    --
    ($(BobPos)+(-0.4,0)$);

\draw[eavesdrop]
    ($(EvePos)+(0,0.7)$) to[bend right=8] (Channel);

\node[agent] (Alice) at (AlicePos)
{
    \includegraphics[
        height=1.65cm
    ]{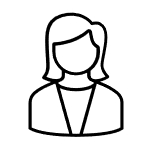}
};

\node[
    font=\small,
    below=2pt of Alice
]
{Alice};

\node[agent] (Bob) at (BobPos)
{
    \includegraphics[
        height=1.65cm
    ]{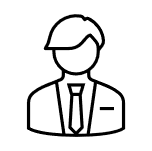}
};

\node[
    font=\small,
    below=2pt of Bob
]
{Bob};

\node[agent] (Eve) at (EvePos)
{
    \includegraphics[
        height=1.65cm
    ]{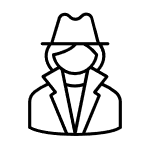}
};

\node[
    font=\small,
    below=2pt of Eve
]
{Eve};

\node[
    inner sep=0pt
]
at ($(Channel)+(0,0.7)$)
{
    \includegraphics[
        width=1.15cm
    ]{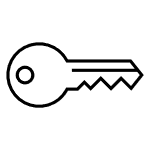}
};

\end{tikzpicture}

\caption{
Secure communication between Alice and Bob.
Alice sends information to Bob using a shared cryptographic key, while
Eve attempts to eavesdrop on the communication channel.
}
\label{Figure: Cryptography}

\end{figure}
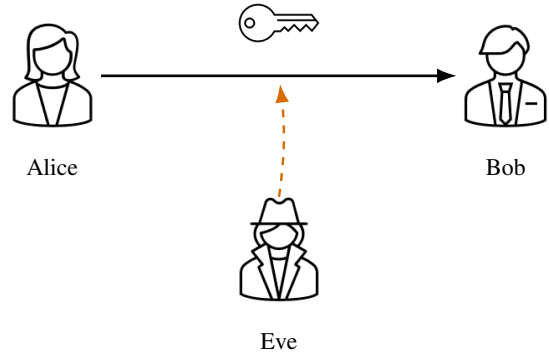

The goal of cryptography is to enable users to communicate and perform information-processing tasks securely, even in the presence of untrusted or adversarial parties, as illustrated in Figure~\ref{Figure: Cryptography}.
Coordination is also closely related to cryptographic primitives such as bit commitment \cite{blum1983coin}. 
Noisy correlations constitute a valuable resource for information-theoretically secure primitives that are otherwise impossible using noiseless communication alone. 
For example, noisy channel simulation  enables commitment protocols \cite{crepeau2020commitment}, secure multi-party computation \cite{goldreich2019play,chaum1988multiparty}, contract signing \cite{even1985randomized}, and zero-knowledge proofs \cite{goldreich1991proofs,brassard1988minimum}. 
One can picture bit commitment as follows. Suppose that 
Alice would like to place a message in a locked box and send the box to Bob, while revealing the key only at a later stage.
Specifically, in a bit-commitment protocol, Alice commits to a bit value in such a way that the value remains hidden from Bob until a later stage, and yet she is prevented from changing the committed value afterward \cite{blum1983coin,crepeau2020commitment}. 
These two requirements are known as the \emph{hiding} and \emph{binding} properties of the protocol. 
Without additional assumptions, perfectly secure bit commitment is impossible using noiseless classical or quantum communication alone \cite{lo1997quantum,mayers1997unconditionally,lo1998quantum}. 
However, noisy channels and imperfect correlations enable information-theoretically secure commitment schemes \cite{winter2003commitment}. 
From a coordination perspective, users seek to generate nontrivial correlations that are strong enough to guarantee security while remaining inaccessible to an adversary. 
Thus, coordination and channel simulation may be viewed as preparation procedures that establish the correlation resources required for secure cryptographic tasks.

\subsubsection{Device independence}

Quantum key distribution (QKD) allows two legitimate users, Alice and Bob,
to establish a shared secret key in the presence of an eavesdropper, Eve
\cite{bennett1984quantum,ekert1991quantum}.
After the protocol, Alice and Bob possess matching secret bits,
while Eve has no information about them.

In a standard QKD protocol, the security proof assumes that Alice's and
Bob's devices operate according to a known mathematical model.
For example, the proof may assume that a source prepares particular
quantum states, and that a measurement device performs the intended
measurements.
If the actual devices behave differently from this model, because of
imperfections or malicious manipulation, these assumptions are no longer valid.

In device-independent quantum key distribution (DI-QKD), the goal is to eliminate the need to trust the internal operation of the devices
\cite{acin2007device,pironio2009device}.
The devices are instead treated as black boxes: Alice and Bob select
measurement inputs and observe the corresponding outputs, without requiring
a detailed description of how the outputs were produced internally.
Security is then inferred from the statistical correlations between their
inputs and outputs.

The central test is the violation of a Bell inequality \cite{bell1964einstein}.
A Bell inequality places an upper bound on the amount of correlation
that can be produced by separated users using only local classical
strategies and pre-shared randomness.
Therefore, if the observed correlations exceed this classical bound,
they cannot be explained by such a local classical model.
The violation provides evidence that Alice and Bob genuinely share  quantum
correlations and also limits how strongly Eve can be correlated with
their outcomes.

If the Bell violation is sufficiently strong and the observed noise is
sufficiently low, Alice and Bob can process their measurement outcomes to
obtain a shared secret key.
From the coordination viewpoint, entanglement provides a resource that
allows Alice and Bob to generate correlations that are simultaneously
strong enough to coordinate their outcomes and sufficiently private from
Eve.
DI-QKD is therefore a coordination-based quantum protocol in which security
is certified from the observed behavior of the distributed devices, rather
than from a complete model of their internal implementation
\cite{devetak2005distillation,wolf2021quantum}.

\subsubsection{Distributed computing}
Coordination arises in distributed computing systems, where multiple users or processors possess only partial local information and must cooperate in order to perform a global computational task. 
Furthermore, the study of coordination is closely connected to a broad range of problems in 
quantum information processing. 
Examples include channel simulation,  %
state merging and redistribution, 
entanglement dilution and distillation, %
distributed compression and source coding,%
randomness extraction, 
channel resolvability and soft covering,
multipartite entanglement manipulation, %
and distributed quantum state transformation. %
Many of these tasks may be viewed as different manifestations of a common underlying problem: the generation or simulation of distributed correlations under communication constraints.

\subsection{Strong vs. Empirical Coordination}
\label{Subsection: Strong vs. Empirical Coordination}

The classical coordination problem was originally introduced by Wyner \cite{wyner1975common}, and later developed in further generality by Cuff \emph{et al.}
\cite{cuff2008communication,cuff2010coordination}.
Unlike conventional communication  tasks, 
 coordination does not aim to merely communicate messages  %
 but rather to create  correlation \cite{%
 le2017joint}.
In what follows, we describe two main notions of coordination.

\paragraph{Strong coordination} Distributed users seek to generate systems whose joint behavior resembles that of a memoryless source, i.e., a source that produces an independent and identically distributed (i.i.d.) sequence. %
This formulation requires simulating the desired correlation across the entire sequence. %

\paragraph{Empirical coordination}
A weaker notion of coordination requires the statistical average behavior of the generated systems to match the target correlation. 
Empirical coordination captures scenarios in which only the observed frequencies or empirical averages are relevant. %

Quantum information theory provides further motivation for both notions of coordination. 
Strong coordination is closely related to tasks aimed at reproducing a prescribed quantum correlation structure across multiple systems. 
Empirical coordination becomes relevant when only the observable statistics of a quantum system are important. 
Since quantum mechanics fundamentally predicts probabilities of measurement outcomes rather than deterministic realizations, many operational tasks depend only on the empirical frequencies generated by repeated experiments \cite{fuchs1996quantum,fuchs2000quantum,bricmont2016making}. 
Examples include distributed sensing, nonlocal games, and quantum data compression, where performance is determined by the observed correlations rather than the exact realization of the underlying quantum state. 
Indeed, one may argue that quantum theory should be viewed primarily as a theory of statistical correlations, making empirical coordination a suitable framework for the study of quantum systems.

The kind of correlation that is to be simulated has a profound impact on the resources required for coordination. 
If the users wish to simulate a probabilistic mixture (i.e., a separable correlation), then classical resources alone are sufficient.
If, however, the users wish to simulate entanglement, such a correlation cannot be reproduced classically. 
Although both tasks can be formulated within the same coordination framework, the communication and entanglement resources required for their implementation may differ substantially.
This distinction highlights a fundamental feature of quantum coordination. %
Specifically, entangled states exhibit correlations that have no classical counterpart and lead to operational advantages in communication, computation, cryptography, and nonlocal games.

\begin{figure}[b]
\caption{Network coordination with 
$K=7$ nodes. The double arrow links are classical while the single arrow ones are quantum. The links are rate limited. For example, Node~2 can send bits to Node~3 at a rate $R_{2 \to 3}$, while, the  qubits rate from Node~6 to Node~5 is $Q_{6 \to 5}$. 
}
\center
\includegraphics[scale=0.85,trim={5.3cm 0 5.5cm 0}]
{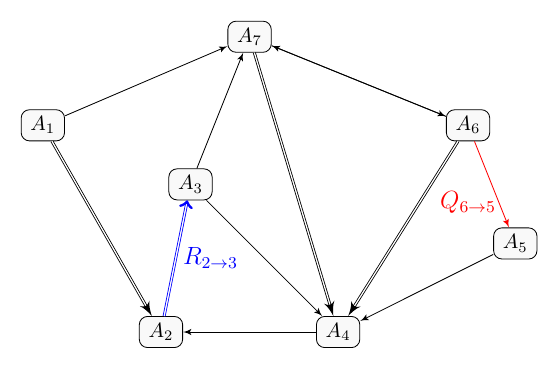} 
\label{Figure: Coordinaiton network}
\end{figure}

We next present the general setting for network coordination. Consider a network consisting of $K$ nodes, where each Node~$i$ is associated with a physical system $A_i$ and performs a local encoding operation $\mathcal{E}_i$. We denote the collection of all network systems by
\begin{align}
\mathbf{A}\triangleq (A_1,\ldots,A_K).
\end{align}
An example of a network with $K=7$ nodes is shown in Figure~\ref{Figure: Coordinaiton network}.
The users are provided with limited communication and shared resources.
Specifically, communication between the nodes occurs over either classical or quantum links. A classical communication link from Node~$i$ to Node~$j$ operates at a rate $R_{i\to j}$ bits per transmission, whereas a quantum communication link operates at a rate $Q_{i\to j}$ of quantum bits (qubits) per transmission.
Furthermore,  entanglement may be shared before communication. We denote the corresponding entanglement rate by $E_{i , j}$. We denote classical and quantum messages by $m_{i\to j}$ and $M_{i\to j}$ respectively. 

\begin{figure}[t]
    \centering
    \begin{minipage}[b]{0.45\textwidth}
        \centering
        \includegraphics[scale=0.75,trim={5.5cm 0 5cm 0}]{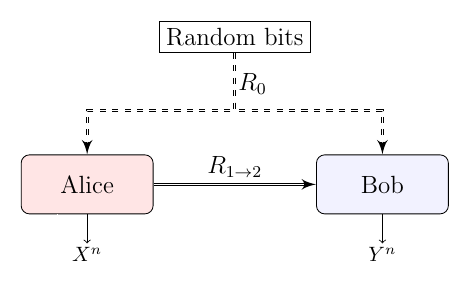}
    
\caption{A two-node classical link and pre-shared common randomness. 
}
        \label{Group_Classical_two_nodes_network}
    \end{minipage}
    \hfill
    \begin{minipage}[b]{0.45\textwidth}
        \centering
        \includegraphics[scale=0.75,trim={7cm 0 7cm -0.5cm}]{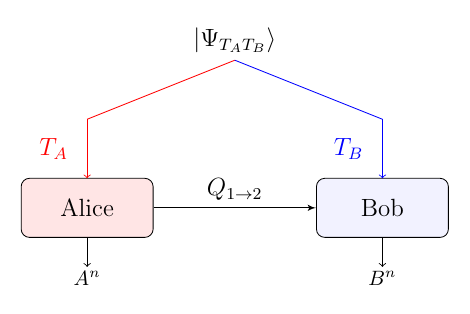}
       \caption{A two-node quantum link pre-shared entanglement.%
       }
        \label{Group_Quantum_two_nodes_network}
    \end{minipage}  
\end{figure}

The network objective is to generate a prescribed correlation among the nodes. This desired correlation is represented by a target joint state
\begin{align}
\omega_{\mathbf{A}},
\end{align}
which the users seek to simulate using their resources. Depending on the coordination criterion, the simulation task can take two different forms. In strong coordination, the users aim to reproduce the statistical behavior of a memoryless source, i.e., to generate a 
joint state that is 
close to $\omega_{\mathbf{A}}^{\otimes n}$. Whereas empirical coordination only guarantees that the average state 
is consistent with 
$\omega_{\mathbf{A}}$.
The basic two-node classical network and quantum network are illustrated in Figures~\ref{Group_Classical_two_nodes_network} and \ref{Group_Quantum_two_nodes_network}, respectively. In the classical setting, Alice and Bob use a classical communication link with rate $R_{1\to 2}$ and possibly pre-shared randomness to generate correlated actions $X^n$ and $Y^n$. In the quantum setting, Alice and Bob seek to generate a joint quantum state by transmitting qubits at a quantum rate $Q_{1\to 2}$ and, if available, exploiting pre-shared entanglement. More generally, the coordination problem asks for the communication and entanglement resources necessary and sufficient to simulate the desired state $\omega_{\mathbf{A}}$ in an arbitrary multi-user network.

\subsection{Two-Node Coordination}
\label{Subsection: Two-Node Coordination}

\subsubsection{Classical network}
The goal in the classical setting is to generate a correlation described by a joint probability distribution.
Specifically, in a two-node network,
 two users would like to simulate a joint distribution $p_{XY}$. Strong coordination can then be achieved with \emph{classical} communication
 if and only if the  rate $R_{1\to2}$ is above  Wyner's common information:
\begin{align}
C(X;Y) \triangleq \min I(U;XY) \,,
\end{align}
where the cost function
$I(U;XY)$ is the mutual information between $U$ and  $(X,Y)$. The minimization is performed over the set of all auxiliary random variables $U$ such that the sequence $X - U - Y$ is a Markov chain.
%
Classical correlation resources
can also be shared before communication takes place. In other words, the users are provided with a shared string of random bits, also known as 
\emph{common randomness} (CR). See Figure~\ref{Group_Classical_two_nodes_network}. 
With sufficient pre-shared CR, the strong coordination requirement is characterized by the 
mutual information between the actions.  
Specifically,
the desired distribution
can be simulated for a classical communication rate of
$R_{1 \to 2}\geq I(X;Y)$ \cite{cuff2008communication,cuff2010coordination}, \cite{bennett2002entanglement}.
On the other hand, in empirical coordination, 
the minimal communication rate is characterized by 
$
R_{1 \to 2}\geq I(X;Y)
$, 
whether CR is available or not \cite{cuff2008communication,cuff2010coordination}. The rate results for the two-node setting are summarized in Figure~\ref{Figure: Two-node table}.

The comparison between the results highlights a fundamental difference between strong and empirical coordination. 
Strong coordination aims to simulate an i.i.d. source.
This requires the entire sequence of actions to be statistically indistinguishable from a realization of a target
i.i.d.  distribution. Strong coordination generally necessitates a larger communication rate, characterized by Wyner's common information \cite{wyner1975common}. 
Empirical coordination imposes a weaker requirement, namely that only the long-term frequencies of the generated actions match the target distribution, leading to the lower rate $I(X;Y)$ \cite{cuff2008communication,cuff2010coordination}. 
As a consequence, CR plays a crucial role in strong coordination, as it reduces the communication rate from Wyner's common information to the mutual information. Nonetheless, in empirical coordination, CR does not 
improve the rate consumption
beyond the mutual information. 
This distinction illustrates the additional resources required when one seeks to reproduce the full statistical behavior of a source rather than only its empirical averages.


\subsubsection{Quantum network}
In a quantum-linked two-node network,  we aim to create a correlation asymptotically indistinguishable from a desired joint quantum state.
The simulation of a bipartite pure state $\ket{\omega}_{AB}$ requires a quantum communication rate of at least the von Neumann entropy of the reduced state 
of  system $B$ \cite{kumagai2016second}, i.e.,  
  $
 Q_{1 \to 2}\geq H(\omega_B) 
 $, 
 with $H(\rho)=-\trace(\rho\log(\rho))$.
 %
In contrast to the classical case, the optimal strategy for Alice is lossless compression of her source \cite{schumacher1995quantum}.

On the other hand, empirical coordination imposes a significantly weaker requirement. 
Rather than reproducing the full joint behavior of a memoryless quantum source, it only requires the empirical average state to converge to the desired target state.
Remarkably, it was recently shown in \cite{natur2025empirical_ITW} that in the quantum setting,
this relaxation does not reduce the communication cost.  
That is, for empirical coordination the optimal rate is also the entropy $H(\omega_B)$.
Consequently, lossless compression remains optimal even under the weaker empirical coordination criterion. 
This contrasts with the classical setting, where empirical coordination requires a lower communication rate. 
In other words, 
in the quantum setting, faithfully reproducing the local statistics of a bipartite state is already as demanding as reproducing the full correlation structure.
Suppose that before communication begins, the nodes have access to pre-shared entanglement resources. See Figure~\ref{Group_Quantum_two_nodes_network}.
Based on the quantum  reverse Shannon theorem \cite{bennett2014quantum},
given 
sufficient 
entanglement,   
the desired state
can be simulated if and only if the quantum communication rate 
is at least equal to one half of the quantum mutual information between the quantum systems of the nodes, i.e., systems $A$ and $B$.
 
\begin{figure}[t]
    \centering
     \renewcommand{\arraystretch}{1.1}

    \begin{tabular}{|c|c|c|}
        \hline
        & Strong Coordination
        & Empirical Coordination
        \\
        \hline
        CR
        & $C(X;Y)$
        & $I(X;Y)$
        \\
        \hline
        No CR
        & $I(X;Y)$
        & $I(X;Y)$
        \\
        \hline
    \end{tabular}
    
    \caption{Comparison of the communication rates in a classically linked two-node network, required for strong and empirical coordination, with and without common randomness.}
    \label{Figure: Two-node table}
\end{figure}

Quantum coordination unifies a wide range of quantum problems; we list a few closely related ones below.

\begin{enumerate}[1.]
    \item 
    \emph{Schumacher compression:} 
    The task requires Alice to losslessly compress a source described by a state $\omega_B^{\otimes n}$. Schumacher’s compression theorem states that the task requires Alice to send qubits to Bob at a rate of at least the entropy of system $B$ \cite{schumacher1995quantum}.
    \item \emph{Channel resolvability:} Given a classical--quantum (c-q) channel, in 
channel resolvability, we seek to approximate the output state of the channel by uniformly selecting an input codeword from a randomly generated codebook and passing it through the channel \cite{hayashi2016quantum}.
Thus, it is equivalent to a c-q state simulation, $\omega_{XB}$, using classical communication.
To achieve this, the communication rate must be at least the mutual information between the systems $X$ and $B$.

    \item 
    \emph{Entanglement dilution:} 
     Alice and Bob seek to transform their pre-shared maximally entangled resource into a bipartite state $\ket{\omega}_{AB}$ using only LOCC: local operations and classical communication.
    Entanglement dilution is possible if the dimension of the maximally entangled state is exponential in the entropy of $\omega_B$, and the required classical communication is negligible 
     \cite{hayden2003communication,harrow2004tight}. 

    \item 
    \emph{State merging:} Assume that  Alice and Bob share  a mixed state $\omega_{AB}$.  %
    In this protocol, 
    Alice 
    sends  
    to Bob her part of the state via classical communication at a rate of at least the mutual information between $A$ and $B$, and quantum communication at a rate of the conditional entropy of $A$ given $B$. 
    \cite{horodecki2005partial,horodecki2007quantum}.
    If the conditional entropy is negative, then the task is achievable with only LOCC, and quantum communication is not needed. 

    \item 
    \emph{State redistribution:}
Alice and Bob share a joint state
 $\ket{\omega_{A B GR}}$, with parts $A$ and $B$ being held by Alice, while Bob has system  $G$, the remaining system $R$ is a  reference.  Alice wants to transfer to Bob part $B$
\cite{devetak2008exact,Yard_Devetak_2009,luo2009channel}. 
State redistribution unifies multiple protocols, for example,  
state merging \cite{horodecki2005partial,horodecki2007quantum}, 
splitting \cite{abeyesinghe2009mother}, and  Schumacher's compression \cite{schumacher1995quantum}. 
   \item 
    \emph{Channel simulation:} 
    In the classical scenario of the reverse Shannon theorem \cite{bennett2002entanglement}, provided with sufficient CR, to simulate any classical channel with a capacity $C$, users must communicate at a bit rate of at least the capacity
 \cite{cuff2008communication,cuff2010coordination}. %
    The quantum analog of this theorem holds  for a product input state, 
    given sufficient pre-shared entanglement, but is not true for a general input 
        \cite{bennett2014quantum}.
\end{enumerate}

The list above is
far from exhaustive. %
Recent years have seen an extensive study of the multi-user versions of the protocols above \cite{abeyesinghe2009mother,ahn2006distributed,9039682,salek2018quantum,Faithful_simulation_Heidari_2019}. 
  Multipartite state merging and splitting have recently been studied by
  Streltsov et al. \cite{streltsov2020rates} and
  George and Cheng~\cite{george2024coherent}, respectively. 
Analysis of simulating the broadcast and multiple-access channels is presented in \cite{cheng2023quantum,cao2024channel,nema2024one}. 

\subsection{General Coordination Problem}
\label{Subsection: General Coordination Problem}

Consider a network of $K$ nodes, with a common goal to create a joint correlation described by a joint quantum state $\omega_{\mathbf{A}}$, while using limited communication and pre-shared correlation resources. 
In the coordination task, we are interested in identifying the optimal trade-offs between the communication and correlation resources. Each user in the network is represented by a node and is assigned an index $i\in[K]$. Each node possesses a block of length $n$ of quantum systems $A_i(1),\ldots,A_i(n)$.

Quantum and classical communication links are available. The rate of the classical link connecting Node~$i$ to Node~$j$ is $R_{i\to j}$, and the corresponding message is $m_{i\to j}$. Similarly, the rate of the quantum link connecting Node~$i$ to Node~$j$ is $Q_{i\to j}$, and the corresponding message is $M_{i\to j}$. The users may also have access to pre-shared correlation resources such as CR with a rate $R_0$, and entanglement assistance with a rate $E_{i,j}$. 
An encoding operation is applied locally by each user on his system  $A_i$, preparing a state $\widehat{\rho}_{\mathbf{A}}$. 

\paragraph{Strong coordination}
Here, the users would like their encoding to simulate a memoryless (i.i.d.) quantum source.
That is, they would like to produce a joint state $\widehat{\rho}_{\mathbf{A}}$ that resembles that of the $n$-fold product state, $\omega_{\mathbf{A}}^{\otimes n}$.  
A rate tuple
\begin{align}
\left(
R_0,
\{R_{i\to j}\},
\{E_{i,j}\},
\{Q_{i\to j}\}
\right)_{i,j\in[K]}
\end{align}
is called achievable if there exists a sequence of codes of length $n=1,2,3,\ldots$, such that the simulation error
\begin{align}
\epsilon_{\text{strong}}(n)=\norm{\widehat{\rho}_{ \mathbf{A}(1)\cdots\mathbf{A}(n)}-\omega_{\mathbf{A}}^{\otimes n}}_1
\end{align}
vanishes as $n$ tends to infinity. 

Classical strong coordination requires that a statistician be unable to reliably distinguish the users' generated action sequence from an i.i.d. sequence drawn from the desired distribution \cite{cuff2008communication,cuff2010coordination}. 
More precisely, if the target behavior is described by a distribution $\pi$, then the induced joint distribution must be statistically indistinguishable from the i.i.d. source $\pi^{\otimes n}$. 
Strong coordination is therefore achieved if there exists a sequence of codes such that
\begin{align}
    \lim_{n\to\infty}
    \norm{
    p_{\mathbf{X}(1)\cdots \mathbf{X}(n)}
    -
    \pi^{\otimes n}
    }_1
    =0.
\end{align}
In other words, not only must the marginal frequencies be correct, but the entire sequence of actions must exhibit the same statistical behavior as a memoryless source. 
This requirement is closely related to channel resolvability and soft covering, where one seeks to synthesize a target distribution rather than merely reproduce its empirical statistics \cite{cuff2013distributed,hayashi2016quantum}.

The quantum formulation is considerably richer than its classical counterpart. 
A quantum state contains both classical correlations and genuinely quantum correlations arising from superposition and entanglement. 
Consequently, strong quantum coordination requires simulating the complete correlation structure of the desired state, including all entanglement and coherence properties. 
This requirement connects strong coordination to a broad family of quantum information-processing tasks, including Schumacher compression, channel simulation, state merging and redistribution, entanglement dilution and distillation, and the quantum reverse Shannon theorem. 
From this perspective, strong coordination is a unifying framework for simulating and manipulating distributed quantum correlations.

\paragraph{Empirical coordination}
In this type of coordination, the criterion is a less stringent condition than strong coordination.
 It requires the convergence of the  \emph{empirical-average} state 
  to the desired correlation $\omega_{\mathbf{A}}$ \cite{cuff2008communication,cuff2010coordination}. 
In empirical coordination, a rate tuple is achievable if there exists a sequence of codes such that the simulation error
\begin{align}
\epsilon_{\text{empirical}}(n)= \norm{\frac{1}{n}\sum_{i=1}^n\rho_{\textbf{A}(i) } -\omega_{\mathbf{A}}}_1
    \label{empirical_quantum}
\end{align}
converges to zero in probability. 

At this point, 
it is only natural for the reader to wonder why we have defined the empirical criterion in this manner. 
We now explain the physical interpretation justifying our criterion for quantum empirical coordination. 

\subsection{Quantum Measurements}
\label{Subsection: Quantum Measurements}

Quantum mechanics emphasizes statistical averages, as it enables calculating probabilities of experimental outcomes rather than describing individual events. For example, according to the uncertainty principle of Heisenberg, the uncertainties in position and momentum cannot both be minimized at the same time \cite{fuchs1996quantum}.
Some scholars, such as Fuchs and Peres \cite{fuchs2000quantum}, contend that quantum theory is confined to representing statistical correlations and does not describe physical reality at all \cite{bricmont2016making}. This approach to quantum mechanics makes empirical coordination very essential for quantum multi-user networks.
Further elaboration is provided below. 

In quantum mechanics, a measurable physical quantity is described by a Hermitian operator $\hat{O}$ in a Hilbert space $\mathcal{H}_{\mathbf{A}}$. Practically, to be able to gather statistics, one has to perform measurements on $n$ systems $(\mathbf{A}(i) \,:\; i\in [n])$.
Consider the $i$th measurement, the expected value with respect to a state $\omega_{\mathbf{A}}$ is
\begin{align}
\langle \hat{O} \rangle_i&=
\trace\left[ \hat{O}\cdot \rho_{\mathbf{A}(i)} \right]
\end{align}
where $i\in [n]$.
Consequently, the empirical average is given by
\begin{align}
\frac{1}{n}\sum_{i=1}^n\langle \hat{O} \rangle_i&=
\trace\left[ \hat{O} \cdot \left( \frac{1}{n}\sum_{i=1}^n \rho_{\mathbf{A}(i)} \right) \right]
\nonumber\\
&=
\trace\left[ \hat{O}\cdot   \overline{\rho}_{\mathbf{A}}  \right]
\end{align}
where we use the notation of $\overline{\rho}_{\mathbf{A}}=\frac{1}{n}\sum_{i=1}^n \rho_{\mathbf{A}(i)}$ for the empirical average state. Notice that different quantum states yield different statistics.
Our criterion ensures that the average statistics of the observed measurement outcomes are close to those induced by the desired state, i.e., 
\begin{align}
\frac{1}{n}\sum_{i=1}^n\langle \hat{O} \rangle_i
&\approx
\trace\left[ \hat{O}\cdot   \omega_{\mathbf{A} } \right] 
\,,
\end{align}
 with high probability.

Similarly, consider a generalized measurement $\{D_\ell: \ell\in [L]\}$ defined in the Hilbert space  $\mathcal{H}_{\mathbf{A}}$. In this type of measurement, each operator corresponds to a specific result. The operators do not necessarily represent a physical quantity; each operator can, for example, indicate the occurrence of an event.
 In the $i$th measurement, the probability for obtaining the outcome $\ell$ is 
$%
p_i(\ell)=\trace(D_\ell^\dagger D_\ell \cdot \rho_{\mathbf{A}(i)} )
$. %
Thereby, the average distribution is  given by 
\begin{align}
\bar{p}(\ell)=\trace(D_\ell^\dagger D_\ell \cdot \overline{\rho}_{\mathbf{A}})
\,.
\end{align}
Thus, empirical coordination guarantees the correct observable behavior of the system, whereas strong coordination guarantees the much stronger property that the entire sequence of quantum systems behaves as if it were generated by a memoryless source. 

\subsection{Outline}
\label{Subsection: Outline}

The purpose of this paper is to provide a unified review of quantum coordination and its role in generating and characterizing quantum correlations, with emphasis on applications to nonlocal games, device-independent quantum key distribution (DI-QKD), and quantum repeaters. In this paper, we focus on illustrating the distinction between classical and quantum resources, the effect of different topologies on the attainable correlations, and the unique phenomena related to the quantum nature of the systems.

The paper has the following structure. We begin in Section~\ref{Section: Notation} by fixing the notation adopted throughout the paper. Section~\ref{Section: Nonlocal Games} then introduces nonlocal games and quantum correlations and explains how they motivate the study of coordination across distributed quantum systems. We discuss nonlocal correlations in Subsection~\ref{Subsection: Nonlocal Correlations}, introduce refereed games in Subsection~\ref{Subsection: Refereed Games}, and review the resources, strategies, and no-signaling constraints that determine the game performance.
Furthermore, we discuss sequential games in Subsection~\ref{Subsection: Sequential Game} and explain how coordination serves as a component of game strategies in Subsection~\ref{Subsection: Coordination as Part of a Game Strategy}.

Section~\ref{Section: Coordination Techniques} introduces the principal coordination techniques that are used in analyzing the information-theoretic models discussed later in the paper. Building on these tools, Section~\ref{Section: Information Theoretic Models: Strong Coordination} reviews strong coordination in quantum networks. We discuss entanglement coordination over quantum links in Subsection~\ref{Subsection: Quantum Links: Entanglement Coordination}, including the broadcast, cascade, and multiple-access networks. We then consider separable correlations over classical links in Subsection~\ref{Subsection: Classical Links: Separable Correlations}, including the broadcast and no-communication networks. These coordination models provide operational insights into the generation of distributed correlations and the resources required to realize quantum and classical game strategies.

Section~\ref{Section: Information Theoretic Models: Empirical Coordination} discusses empirical coordination and its quantum extensions. We consider coordination with quantum links and entanglement assistance in Subsection~\ref{Quantum Links: Empirical Entanglement Coordination}, and coordination with classical links and separable correlations in Subsection~\ref{Subsection: Classical Links: Empirical Separable Correlations}. Particular attention is given to the distinction between strong and empirical notions of coordination and their relevance to distributed tasks and nonlocal games.

Finally, Section~\ref{Section: Discussion and Conclusions} discusses broader implications and applications of quantum coordination. In particular, we examine connections to DI-QKD and quantum repeaters in Subsections~\ref{Subsection: Device Independent QKD} and \ref{Subsection: Quantum Repeaters}, respectively. We conclude in Subsection~\ref{Subsection: Summary and Future Directions} by discussing open problems and future directions at the intersection of coordination theory, quantum networks, and nonlocal games.

\section{Notation and Basic Definitions}
\label{Section: Notation}
Below we state the notation used throughout the paper. Capital letters $X,Y,\ldots$ are used for discrete random variables over alphabets $\mathcal{X},\mathcal{Y},\ldots$ of finite size. A random variable $X$ has a probability mass function (PMF) on $\mathcal{X}$ denoted by $p_X(x)$. 
  The statistical distinguishability of the PMFs $p_X$ and $q_X$ is measured by the total variation distance, which is given by
\begin{align}
\frac{1}{2}\norm{p_X-q_X}_1 =\frac{1}{2}\sum_{a\in \mathcal{X}}\abs{p_X(a)-q_X(a)} .
\end{align}

For $X\sim p_X$, the Shannon entropy is
$
H(p_X)
=
\sum_{x\in\operatorname{supp}(p_X)}
p_X(x)\log_2\frac{1}{p_X(x)}.
$
We also write $H(X)\equiv H(p_X)$. For a joint PMF $p_{XY}$, we similarly write $H(XY)\equiv H(p_{XY})$.
For jointly distributed random variables $X$ and $Y$, define the mutual information between them as
$
I(X;Y)=
H(X)+H(Y)-H(XY).
$
The conditional entropy with respect to 
$p_X\times p_{Y|X}$
is defined as
$H(Y|X)=\sum_{x\in\mathrm{supp}(p_X)} p_X(x)H(Y|X=x)
$, where $H(Y|X=x)\equiv H(p_{Y|X}(\cdot|x))$.
The binary entropy function is $h_2(x)=-x\log_2(x)-(1-x)\log_2(1-x)$ for $x\in[0,1]$, with the convention $0\log_2(0)=0$.

For a blocklength $n$, let $[n]=\{1,\ldots,n\}$ and denote a sequence over $\mathcal{X}$ by $x^n=(x_1,\ldots,x_n)$. The relative frequency of each symbol $a\in\mathcal{X}$ is represented by the type $\bar{\mathsf{P}}_{x^n}(a)=\frac{1}{n}\sum_{i=1}^n\mathbf{1}\{x_i=a\}$, where $\mathbf{1}\{\cdot\}$ is the indicator function.

Throughout the paper, all quantum systems are represented by finite-dimensional Hilbert spaces. We write $\ket{v}\in\mathcal{H}$ for a vector, $\bra{v}$ for its Hermitian conjugate, and $D^\dagger$ for the adjoint of an operator $D\in L(\mathcal{H})$. For Hermitian operators $P$ and $Q$, their normalized trace distance is
\begin{align}
        \frac{1}{2}\norm{P-Q}_{1}=\frac{1}{2}\Tr\Big[\abs{P-Q}%
        \Big]
\end{align}

A quantum system $A$ is associated with a Hilbert space $\mathcal{H}_A$, and its state is represented by a density operator $\rho_A\in L(\mathcal{H}_A)$, namely, a positive semidefinite operator with unit trace. A generalized measurement with outcomes $j\in[N]$ is described by a collection of positive semidefinite operators $\{D_j:j\in[N]\}$ satisfying $\sum_{j=1}^N D_j=\identity$. This collection is called a positive operator-valued measure (POVM). According to the Born rule, outcome $j$ occurs with probability $p_J(j)=\trace(D_j\rho_A)$.

The von Neumann entropy of $\rho_A$ is
$H(\rho_A)=-\trace[\rho_A\log\rho_A]$, we also write $H(A)_\rho=H(\rho_A)$. Similarly,
$H(AB)_\rho=H(\rho_{AB})$. Pure states have zero entropy. For
$\rho_{AB}$, the conditional entropy 
is $H(A|B)_\rho=H(AB)_\rho-H(B)_\rho$. 
The coherent information is
$I(A\rangle B)_\rho=-H(A|B)_\rho$ and is a central
quantity in quantum communication and entanglement distillation.

A separable bipartite state $\rho_{AB}$ is a state of the following form
\begin{align}
\rho_{AB}
=
\sum_{x\in\mathcal{X}}
p_X(x)\,\rho_x\otimes\sigma_x,
\end{align}
for some PMF $p_X$. Otherwise, it is entangled. Entangled states may have negative conditional entropy, equivalently positive coherent information.

Similarly, a multipartite state is  separable if
\begin{align}
\rho_{A_1\ldots A_K}
=
\sum_{x\in\mathcal{X}}
p_X(x)\,
\rho_x^{(1)}\otimes\cdots\otimes\rho_x^{(K)}.
\end{align}
 
A classical--quantum (c-q) state is written as
\begin{align}
\omega_{XB}
=
\sum_{x\in\mathcal{X}}
p_X(x)\ketbra{x}_X\otimes\omega_B^x,
\end{align}
where $X$ is classical, $p_X$ is its PMF, and $\omega_B^x$ is the quantum state associated with $x$.

A quantum channel $\mathcal{N}_{A\to B}$ maps an input state on $A$ to an output state on $B$ and is represented by a completely positive, trace-preserving map. A c-q channel assigns a quantum state $\omega_B^x=\mathcal{N}_{X\to B}(x)$ for each $x\in\mathcal{X}$.

\section{Nonlocal Games}
\label{Section: Nonlocal Games}

\subsection{Nonlocal Correlations}
\label{Subsection: Nonlocal Correlations}

\begin{figure}[t]
    \centering
    \hspace{0.5cm}
    \begin{minipage}[b]{0.3\linewidth} 
        \centering
        \includegraphics[scale=0.7,trim={5.3cm 0 5.5cm 0}]
        {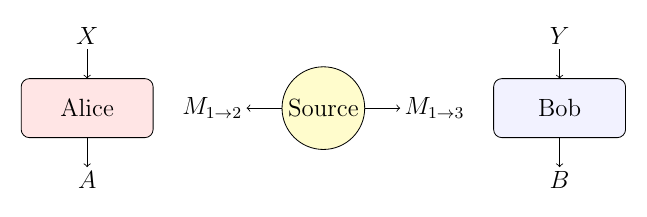} %
        \begin{center}
        \footnotesize{(a) Bell experiment setting.}
        \end{center}
    \end{minipage}
    \hspace{3.5cm}
    \begin{minipage}[b]{0.3\linewidth} 
        \centering
         \includegraphics[scale=0.7,trim={5.3cm -1cm 5.5cm 0}]
        {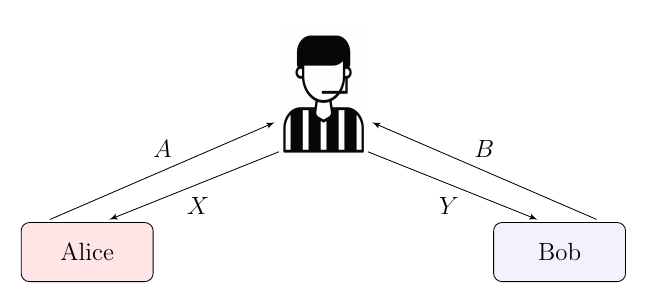} %
        \begin{center}
        \footnotesize{(b) Refereed game setup.}
        \end{center}
    \end{minipage}
    \caption{From Bell Experiments to Refereed Games. Figure~(a) provides a typical description of a Bell experiment setting; the Source prepares the physical systems $M_{1\to2}$ and $M_{1\to3}$ and sends them to the observers Alice and Bob, respectively. After receiving the systems, each of them chooses a measurement device, $X$ for Alice and $Y$ for Bob. The corresponding classical outcomes are $A$ and $B$. Figure~(b) is the game setup of two players and a referee. Alice and Bob receive their questions $X$ and $Y$ from the referee, and send their answers $A$ and $B$ to the referee, respectively. If the tuple $(X,Y,A,B)=(x,y,a,b)$ satisfies the game condition, then the round is won.}
     \label{Figure: Bell experiments and refereed games}
\end{figure}

Quantum correlations play a central role in nonlocal games and Bell experiments \cite{brunner2014bell}.
In a Bell experiment, 
a source prepares and sends physical systems $M_{1\to2}$ and $M_{1\to3}$ to two parties, typically Alice and Bob. See Figure~\ref{Figure: Bell experiments and refereed games}~(a). 
In this paper, we refer to the elements of the Bell setting as the  Source, 
Alice, and Bob, abbreviated by $S$, $A$, and $B$. Once Alice receives $M_{1\to2}$ and Bob receives $M_{1\to3}$, they each choose a measurement from a given set of measurements. We refer to the index of the chosen measurements by $X$ for Alice and $Y$ for Bob. The classical measurement outcomes are $A$ and $B$, respectively.  

The measurement results  $A$ and $B$ can change from round to round due to the probabilistic nature of quantum mechanics \cite{wilde2017quantum}, even when applying the same measurements $X$ and $Y$. The statistics of the measurements are consistent with a conditional probability distribution denoted by $P_{AB|XY}(a,b|x,y)$, which can be estimated by repeating the experiment sufficiently many times. 
The conditional probability law is also referred to as \emph{correlation} or behavior.
Generally, a behavior is not  necessarily separable, i.e., it is not always possible to write $P_{AB|XY}$  as $
P_{A|X}\times 
P_{B|Y}
$, even when a long distance separates the observers. 
However, the presence of correlations doesn't indicate causal influence between the systems.
Locality means that all common influences from the past can be represented by a random variable $U$, commonly called a hidden variable \cite{einstein1935can,bell1966problem}. Once $U$ is specified, the outcomes become conditionally independent given the respective inputs, so their correlation is entirely explained by the hidden variable, i.e., %
\begin{align}
    P_{AB|XY}(a,b|x,y)=\int_{\text{supp}(U)}  {p_{U}(u)P_{A|XU}(a|x,u)P_{B|YU}(b|y,u) \,\text{d}u}\,.
    \label{Equation: locality}
\end{align}

Quantum theory predicts that, for certain measurements on entangled systems, the resulting correlations violate the locality condition in \eqref{Equation: locality}.

Suppose that Alice and Bob share a bipartite state
$\rho_{M_{1\to2}M_{1\to3}}$. For each pair of questions $(x,y)$,
they perform the local measurements
$\{F_{a|x}:a\in\mathcal{A}\}$ and
$\{D_{b|y}:b\in\mathcal{B}\}$, respectively. The resulting correlation is
\begin{align}
P_{AB|XY}(a,b|x,y)
=
\Tr\!\left[
\left(F_{a|x}\otimes D_{b|y}\right)
\rho_{M_{1\to2}M_{1\to3}}
\right].
\end{align}
The collection of all correlations that can be obtained in this manner
is commonly denoted by $C_q$ \cite{slofstra2019set}.

One of the standard examples of quantum nonlocality is the CHSH experiment, named after Clauser, Horne, Shimony, and Holt \cite{clauser1969proposed}. Consider the Bell experiment illustrated in Figure~\ref{Figure: Bell experiments and refereed games}~(a). Alice and Bob each choose between two measurements, indexed by $X,Y\in\{0,1\}$, and obtain binary outcomes $A,B\in\{+1,-1\}$. Define the CHSH quantity
\begin{align}
T
=
\langle A_0B_0\rangle
+
\langle A_0B_1\rangle
+
\langle A_1B_0\rangle
-
\langle A_1B_1\rangle,
\label{Equation: T quantity}
\end{align}
where, for every $(x,y)\in\mathcal{X}\times\mathcal{Y}$,
$\langle A_xB_y\rangle
=
\sum_{a,b\in\{\pm1\}}
ab\,P_{AB|XY}(a,b|x,y)$
is the correlation between the corresponding outcomes.

If the correlation $P_{AB|XY}$ satisfies the locality condition in
\eqref{Equation: locality}, then the CHSH quantity must satisfy
$\lvert T\rvert\leq 2$. Quantum correlations, however, may violate this
bound. Suppose that Alice and Bob each receive one qubit of the EPR pair
\begin{align}
\ket{\Phi}_{M_{1\to2}M_{1\to3}}
=
\frac{1}{\sqrt{2}}
\left(
\ket{00}+\ket{11}
\right).
\end{align}
Let $(\Sigma_1,\Sigma_2,\Sigma_3)$ denote the Pauli operators. Alice
chooses her measurement according to $X$: for $X=0$, she measures
$\Sigma_3$, whereas for $X=1$, she measures $\Sigma_1$. Bob measures
$\frac{\Sigma_3+\Sigma_1}{\sqrt{2}}$ when $Y=0$ and
$\frac{\Sigma_3-\Sigma_1}{\sqrt{2}}$ when $Y=1$. These measurements yield
$T=2\sqrt{2}>2$. Therefore, the resulting quantum correlation violates
the CHSH inequality and cannot be reproduced by a local hidden-variable
model \cite{bell1964einstein}.

\subsection{Refereed Games}
\label{Subsection: Refereed Games}

Refereed games provide an equivalent way to describe the Bell setting.
Consider the game illustrated in
Figure~\ref{Figure: Bell experiments and refereed games}~(b).
The referee draws a pair of questions $(X,Y)\in\mathcal{X}\times\mathcal{Y}$
according to a distribution $p_{XY}$, and sends $X$ to Alice and $Y$ to
Bob. After receiving their questions, Alice and Bob return classical
answers $A\in\mathcal{A}$ and $B\in\mathcal{B}$, respectively.

The winning rule is specified by a condition $\mathscr{W}$ on the tuple
$(X,Y,A,B)$. Its indicator function is defined as
\begin{align}
V(x,y,a,b)
=
\begin{cases}
1, & \text{if $(x,y,a,b)$ satisfies $\mathscr{W}$},\\
0, & \text{otherwise}.
\end{cases}
\label{Equation: Indicator}
\end{align}

We refer to one execution of this procedure as a single-shot game. We now describe how the game is implemented and the rules that Alice and Bob must follow.

$\,$

\subsubsection{Resources}
As in the Bell setting, a source distributes correlated physical systems to Alice and Bob before the game begins; see Figure~\ref{Figure: Bell experiments and refereed games}~(a). We refer to this source of correlation resources as the \emph{Source} ($S$).

$\,$

\subsubsection{Strategy}
Before the referee chooses the questions, the Source, Alice, and Bob agree on a strategy and the correlation resources needed to apply the strategy. Both the strategy and the required resources depend on the rules of the game.

$\,$

\subsubsection{No Signaling}
During the game, Alice and Bob are not allowed to communicate. They can use the pre-shared correlation resources and perform local quantum measurements to coordinate their answers.

$\,$

The game can equivalently be described in three phases. In Phase~1, the Source distributes the correlation resources $M_{1\to2}$ and $M_{1\to3}$ to Alice and Bob. In Phase~2, the referee draws $(x,y)$ according to $p_{XY}$ and sends $x$ to Alice and $y$ to Bob. In Phase~3, Alice and Bob use their questions and shared resources to produce the answers $A$ and $B$, which are then sent to the referee. The referee uses these answers to determine whether they have won the game.

The winning probability is given by
\begin{align}
\pi(P_{AB|XY})
=
\sum_{(x,y,a,b)\in
\mathcal{X}\times\mathcal{Y}\times\mathcal{A}\times\mathcal{B}}
p_{XY}(x,y)
P_{AB|XY}(a,b|x,y)
V(x,y,a,b).
\label{Equation: Winning probability}
\end{align}
Thus, the performance of the players depends on the correlation
$P_{AB|XY}$ generated through the three phases described above.
For example, in the CHSH game, the players win when
$x\wedge y=a\oplus b$, where $x,y,a,b\in\{0,1\}$. For uniformly distributed questions, the winning probability is
$\pi(P_{AB|XY})=\frac{1}{2}\left(1+\frac{T}{4}\right)$. Classical correlations satisfy $T\leq 2$; see \eqref{Equation: T quantity}. Therefore, any classical strategy has winning probability at most $0.75$. By using entanglement, Alice and Bob can achieve $T=2\sqrt{2}$ and win with probability
$\pi(P_{AB|XY})=\frac{1}{2}\left(1+\frac{1}{\sqrt{2}}\right)\approx0.8535$.

\subsection{Sequential Game}
\label{Subsection: Sequential Game}

\begin{figure}[t]
    \centering
    \begin{minipage}[b]{0.3\linewidth} %
        \centering
        \includegraphics[scale=0.65,trim={5.3cm 0cm 5.5cm 0}]{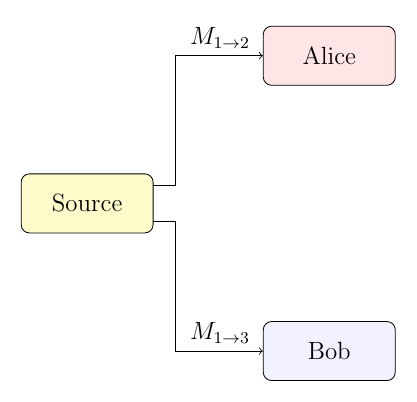} %
        \vspace{0.25cm}
        \begin{center}
        \footnotesize{(a) Phase~1.} %
        \end{center}
    \end{minipage}
    \hspace{0.5cm}%
    \begin{minipage}[b]{0.3\linewidth} %
        \centering
        \includegraphics[scale=0.65,trim={5.3cm 0 5.5cm 0}]{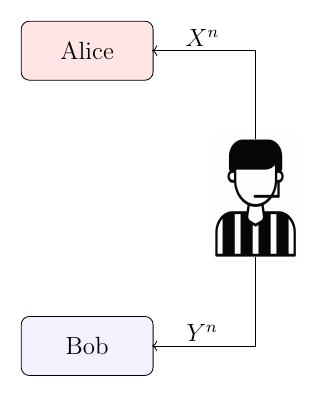} %
        \vspace{0.25cm}
        \begin{center}
        \footnotesize{(b) Phase~2.}%
        \end{center}
    \end{minipage}
    \hspace{0.5cm}%
    \begin{minipage}[b]{0.3\linewidth} %
        \centering
        \includegraphics[scale=0.65,trim={5.3cm 0 5.5cm 0}]{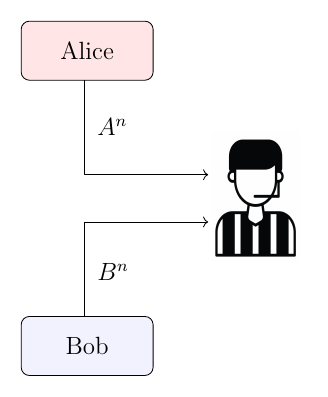} %
        \vspace{0.25cm}
        \begin{center}
        \footnotesize{(c) Phase~3.}%
        \end{center}
    \end{minipage}
    \caption{ Phased implementation of a refereed game:  %
(a) 
In the first phase, Alice and Bob are provided with correlated physical systems prepared by the Source. 
(b) 
In the second phase, Alice and Bob receive the questions $X$ and $Y$ from the referee.
(c)
In the third phase, Alice and Bob send their answers $A$ and $B$ to the referee, who checks whether they won the round. %
}
    \label{Figure: Nonlocal games}
\end{figure}
The implementation of a sequential nonlocal game is illustrated in Figure~\ref{Figure: Nonlocal games}. 
In Phase~1, the correlation resources are distributed between the players by the Source. 
In Phase~2, a sequence of question pairs $(x_i,y_i)$ is generated by the referee independently according to a PMF $p_{XY}$, and the corresponding questions are sent to each player.
In Phase~3, Alice and Bob send their answer sequences $A^n$ and $B^n$ to the referee to check for winning according to a scoring function $V(x,y,a,b)$.

The distinction between strong and empirical coordination arises from the manner in which the correlation is reproduced over multiple rounds of the game.

\paragraph{Strong coordination}

Under strong coordination, the players seek to simulate the entire sequence of game rounds as if it were generated by a memoryless source with correlation $P_{AB|XY}$. Let $q_{X^nY^nA^nB^n}$ denote the induced distribution.
The induced conditional distribution must satisfy
\begin{align}
q_{A^nB^n|X^nY^n}
\approx
P_{AB|XY}^{\otimes n}.
\end{align}
Consequently, every finite collection of game rounds behaves as if it were generated independently according to the target correlation. 
The worst-case winning probability can be characterized by
\begin{align}
\pi_n(q_{A^nB^n|X^nY^n})
=
\min_{i\in[n]}
\sum_{x^n,y^n,a^n,b^n}
p_{XY}^{\otimes n}(x^n,y^n)
q_{A^nB^n|X^nY^n}(a^n,b^n|x^n,y^n)
V(x_i,y_i,a_i,b_i).
\end{align}
If
\begin{align}
q_{A^nB^n|X^nY^n}
\approx
P_{AB|XY}^{\otimes n},
\end{align}
then
\begin{align}
\pi_n(q_{A^nB^n|X^nY^n})
\approx
\pi(P_{AB|XY}).
\end{align}
Strong coordination, therefore, guarantees that the entire game realization is statistically indistinguishable from repeated independent executions of the target strategy.

\paragraph{Empirical coordination}

Empirical coordination imposes a weaker requirement. 
Rather than reproducing the full joint distribution across all game rounds, it only requires the players' empirical correlation to converge to the target correlation. 
 The average winning statistics are given by
\begin{align}
\langle V\rangle
&=
\mathbb{E}
\left[
\frac{1}{n}
\sum_{i=1}^{n}
V(X_i,Y_i,A_i,B_i)
\right]
\nonumber\\
&=
\sum_{x,y,a,b}
p_{XY}(x,y)
\bar q_{AB|XY}(a,b|x,y)
V(x,y,a,b),
\end{align}
where
\begin{align}
\bar q_{AB|XY}(a,b|x,y)
=
\frac{1}{n}
\sum_{i=1}^{n}
q_{A_iB_i|X_iY_i}(a,b|x,y).
\end{align}

If the empirical correlation satisfies
\begin{align}
\bar q_{AB|XY}
\approx
P_{AB|XY},
\end{align}
then
\begin{align}
\langle V\rangle
\approx
\pi(P_{AB|XY}).
\end{align}

Unlike strong coordination, empirical coordination does not require that the game rounds be independent or that the entire sequence resemble a memoryless source. 
Arbitrary dependencies between different rounds are permitted, provided that the empirical correlation converges to the desired one. 
Since Bell experiments and nonlocal games are ultimately evaluated using observed frequencies and expectation values obtained from many repetitions of the experiment, empirical coordination captures the minimal requirement needed to reproduce the asymptotic game statistics. 
Strong coordination, on the other hand, guarantees the stronger property that the entire sequence of plays behaves as if it were generated independently according to the target correlation.

\subsection{Coordination as Part of a Game Strategy}
\label{Subsection: Coordination as Part of a Game Strategy}
In this subsection, we explicitly illustrate the relation between coordination and refereed games.
The broadcast coordination network can be viewed as a preparation stage that generates the correlation resources later used by the players in the game. 
We assume that registers $A$ and $B$ are classical, and $S$ is null (for instance, $\mathrm{dim}(\mathcal{H}_S)=1$).

The game strategy is described by a correlation $P_{AB|XY}$ and the corresponding  state
\begin{align}
\omega_{XYAB}
=
\sum_{(x,y,a,b)}
p_{XY}(x,y)
P_{AB|XY}(a,b|x,y)
\ketbra{x,y,a,b}.
\end{align}
We denote the set of all correlations that achieve a winning probability of at least $\gamma$ by $\mathfrak{S}(\gamma)$.

\paragraph{Strong coordination}

In strong coordination, the players aim to generate a sequence of rounds that behaves as if it were produced by a memoryless source described by $P_{AB|XY}$.
Thus, the induced state satisfies
\begin{align}
\widehat{\rho}_{X^nY^nA^nB^n}
\approx
\omega_{XYAB}^{\otimes n}.
\end{align}
To achieve a winning probability of at least $\gamma$, the source must communicate at rates that enable strong coordination of a correlation $P_{AB|XY}\in\mathfrak{S}(\gamma)$.
\paragraph{Empirical coordination}

Empirical coordination imposes a weaker requirement, 
where the empirical average $\overline{\rho}_{XYAB}$ state must satisfy 
\begin{align}
\overline{\rho}_{XYAB}
\approx
\omega_{XYAB},
\end{align}
where $\overline{\rho}_{XYAB}$ is defined as in Subsection~\ref{Subsection: General Coordination Problem}.
Unlike the strong coordination case, Alice and Bob achieve an average winning probability.

Since Bell experiments and nonlocal games are ultimately evaluated through expectation values obtained from many repetitions of the experiment, empirical coordination emerges naturally.

\subsection{Example: The CHSH game}
\label{Subsection: CHSH}
The CHSH game is a central example used to demonstrate the role of nonlocal correlations in quantum information. Suppose that the Source distributes qubit pairs to Alice and Bob in the state 
\begin{align}
    \ket{\omega^{(x,y)}
    }&=\sqrt{\alpha_{x,y}}\ket{00}+\sqrt{1-\alpha_{x,y}}\ket{11}\,,
\end{align}
where the probability amplitude depends on the classical registers $X$ and $Y$.
If we apply the optimal CHSH strategy shown in Subsection~\ref{Subsection: Nonlocal Correlations}, and denote the obtained correlation by $P$, then the probability of winning is 
\begin{align}
    \pi^{\text{CHSH}}(P)=\frac{1}{16}\sum_{x,y\in \{0,1\}}\left[\frac{1+2\sqrt{2}}{\sqrt{2}}+\sqrt{2\alpha_{x,y}(1-\alpha_{x,y})}\right]\,.
\end{align}
\begin{figure}[t]
    \centering
    \includegraphics[scale=0.22,trim={5.3cm 0 5.5cm 0}]{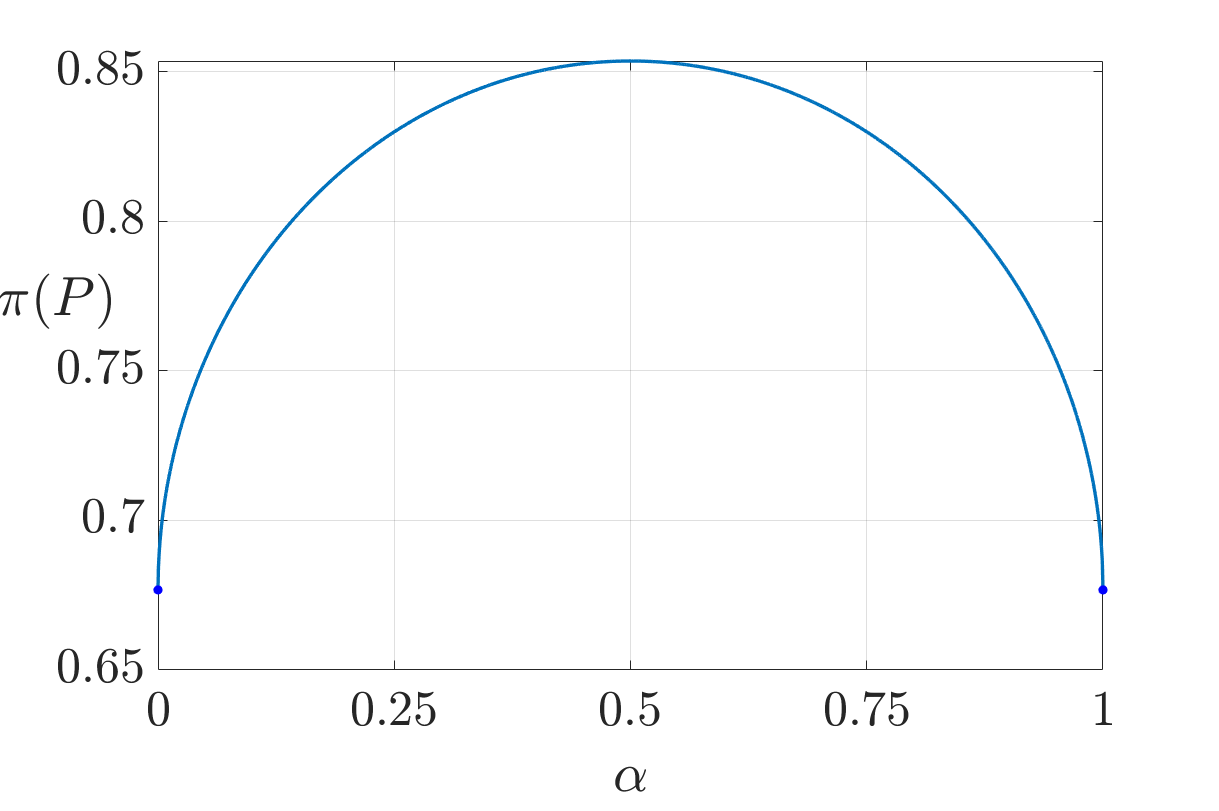}
    \centering
    \caption{ Winning probability as a function of the probability amplitude $\alpha$.}  
    \label{Figure: Winning probability}
\end{figure}
To maximize the Bell violation \cite{cirel1980quantum}, we set
$\alpha_{x,y}=\frac{1}{2}$ for all $(x,y)\in \mathcal{X}\times\mathcal{Y}$, and obtain the maximum winning probability the game $\pi^{\text{CHSH}}(P)
= 0.8535$.
On the other hand, if we set 
$\alpha_{x,y}\equiv 0$, then  no correlation remains, and the winning probability is $\pi^{\text{CHSH}}(P)=0.6767$. Thus, in this case, the quantum CHSH measurement strategy performs even worse than the optimal classical strategy, whose winning probability is $0.75$.

 In both types of coordination, empirical and strong, the required communication rates in Phase~1 are $Q_{1\to 2}\geq 
\frac{1}{2}h_2\left( 
 \frac{1}{2}\left( \alpha_{0,0}+\alpha_{0,1} \right)\right)+\frac{1}{2}h_2\left( 
 \frac{1}{2}\left( \alpha_{1,0}+\alpha_{1,1} \right)\right)$ and $Q_{1\to 3}\geq 
 \frac{1}{2}h_2\left( 
 \frac{1}{2}\left( \alpha_{0,0}+\alpha_{1,0} \right)\right)+\frac{1}{2}h_2\left( 
 \frac{1}{2}\left( \alpha_{0,1}+\alpha_{1,1} \right)\right)$. The function $h_2(\cdot)$ is the binary entropy. The rates become $Q_{i\to j}\geq h_2(\alpha)$ in the special case of $\alpha_{x,y}=\alpha$ for all $(x,y)\in\mathcal{X}\times\mathcal{Y}$.
The CHSH measurement strategy matches the best classical winning probability at a threshold value $\alpha^*$. Above this value,  
we obtain a Bell violation. This happens when $\alpha_{x,y}>0.04491$ for all $(x,y)\in\mathcal{X}\times\mathcal{Y}$.
Figure~\ref{Figure: Winning probability} shows the winning probability in the CHSH game as a function of $\alpha$. 
Notice that near $\alpha=0$, the gradient is steep, meaning a small change in $\alpha$ affects the winning probability significantly. %
However, closer to $\alpha=\frac{1}{2}$, the slope diminishes; thus, small changes in $\alpha$ in this range do not affect the winning probability as much as near the end values $0$ and $1$. 
 To violate the Bell inequality, the required coordination rates are
$Q_{i\to j}\geq h_2(\alpha^*)= 0.2643$.

\begin{remark}
    Notice that in this case the rates for strong and empirical coordination are identical, see Theorems~\ref{Theorem: Broadcast - Strong Quantum links} and \ref{Theorem: Broadcast Empirical} below.
    We will see that for networks with classical links, the strong and empirical coordination rates differ (see Subsections~\ref{Subsubsection: Cascade Network}, \ref{Subsection: Classical Links: Empirical Separable Correlations}).
\end{remark}

The ability of quantum coordination for creating nontrivial correlations is of great significance in other types of games as well. For example, in pseudo-telepathy games, winning with certainty is guaranteed when using quantum strategies.
Another illustration is provided by the magic square game \cite{brassard2005quantum}. The inputs $X$ and $Y$ specify a row and a column of the square, respectively. Each player returns a three-bit string satisfying the required parity constraint, and their answers must agree at the intersecting cell. In this setting, the desired strategy can be implemented using a quantum communication rate of $Q_{i\to j}=2$ qubits per question on each relevant link.
Slofstra and Vidick \cite{slofstra2018entanglement} constructed a game in which a suitable correlation achieves a winning probability of $1-e^{-\Delta}$, while requiring a communication rate that scales as $Q_{i\to j}\propto\Delta$ qubits per round.

In quantum games and Bell measurements, the devices used may be imperfect. Detector inefficiency refers to the situation in which a measurement device fails to produce a conclusive outcome. 
Dilley and Chitambar \cite{dilley2018more} considered an asymmetric Bell experiment in which Alice's detector is assumed to be perfect, while Bob obtains a conclusive measurement outcome only a fraction $\eta$ of the time. 
The remaining fraction $1-\eta$ corresponds to inconclusive measurement events, which modify the observed statistics and affect the achievable violation of the CHSH inequality. 
They showed that, under this operational constraint, weakly entangled states $\omega_{AB}$ can produce larger violations of the CHSH inequality than a maximally entangled state, such as $\ket{\omega_{AB}}=\frac{1}{\sqrt{2}}(\ket{00}+\ket{11})$. 
Thus, the state that maximizes Bell nonlocality in the presence of detector losses is not necessarily the state with the largest amount of entanglement, see Figure~\ref{Figure: Detection}. 
Coordination is particularly relevant in such scenarios because our interest is in the observable correlation generated among the distributed users, rather than the quantum state itself.

\usetikzlibrary{shapes.misc}

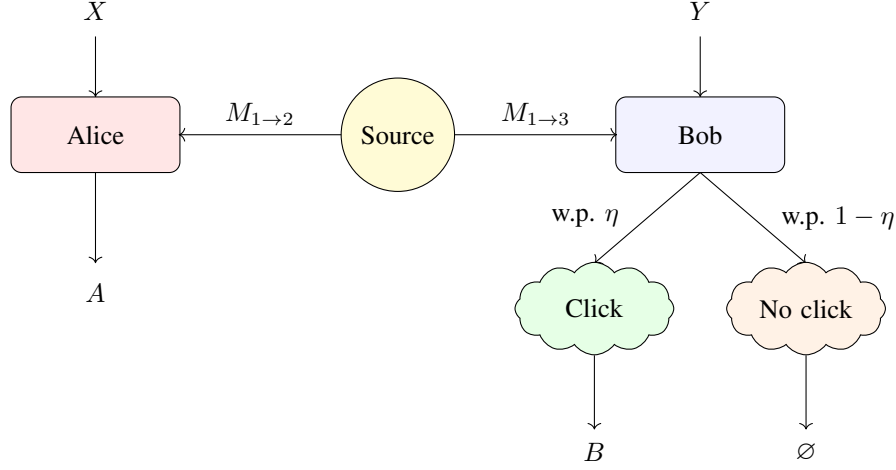
\begin{figure}[t]
\centering

\begin{tikzpicture}[node distance = 1cm, auto]

\tikzset{
    block_1/.style={
        rectangle,
        draw,
        fill=red!10,
        text width=2cm,
        text centered,
        rounded corners,
        minimum height=1cm
    },
    block_2/.style={
        rectangle,
        draw,
        fill=blue!5,
        text width=2cm,
        text centered,
        rounded corners,
        minimum height=1cm
    },
    cloudbase/.style={
    draw,
    cloud,
    cloud puffs=12,
    cloud puff arc=110,
    aspect=2,
    fill=green!10,
    minimum width=2.1cm,
    minimum height=1cm,
    text width=1.4cm,
    text centered,
    align=center,
    inner sep=0pt
},
clickcloud/.style={
    cloudbase,
    fill=green!10
},
noclickcloud/.style={
    cloudbase,
    fill=orange!10
}
}

\node[block_1] (Alice) at (1.5,0.5) {Alice};
\node[block_2] (Bob) at (9.5,0.5) {Bob};

\draw[draw=black,fill=yellow!20]
(5.5,0.5) circle[radius=0.75cm]
node {Source};

\draw[->]
(1.5,1.8) -- (1.5,1.0)
node[midway,above=0.5cm] {$X$};

\draw[->]
(9.5,1.8) -- (9.5,1.0)
node[midway,above=0.5cm] {$Y$};

\draw[->]
(4.75,0.5) -- (2.6,0.5)
node[midway,above] {$M_{1\to2}$};

\draw[->]
(6.25,0.5) -- (8.4,0.5)
node[midway,above] {$M_{1\to3}$};

\draw[->]
(1.5,0) -- (1.5,-1.2)node[midway,below=0.75] {$A$};

\draw[->]
(9.5,0) -- (8.1,-1.2)
node[midway,left=0.25cm] {w.p. $\eta$};

\draw[->]
(9.5,0) -- (10.9,-1.2)
node[midway,right=0.25cm] {w.p. $1-\eta$};

\node[clickcloud] (Bclick) at (8.1,-1.8) {Click};

\node[noclickcloud] (Bnoclick) at (10.9,-1.8) {No click};

\draw[->]
(Bclick) -- (8.1,-3.4);

\draw[->]
(Bnoclick) -- (10.9,-3.4);

\node at (8.1,-3.7) {$B$};

\node at (10.9,-3.7) {$\varnothing$};

\end{tikzpicture}

\caption{
Asymmetric Bell experiment with detector inefficiency.
A source distributes entangled physical systems $M_{1\to 2}$ and $M_{2\to 3}$ to Alice and Bob, respectively.
Alice's detector is assumed to be ideal and always produces a measurement outcome.
Bob's detector produces a conclusive outcome only a fraction $\eta$ of the time, while the remaining fraction $1-\eta$ corresponds to inconclusive measurement events.
The detector losses modify the correlation $P_{AB|XY}$ and affect the violation of the CHSH inequality.
}
\label{Figure: Detection}

\end{figure}

\section{Coordination Techniques}
\label{Section: Coordination Techniques}

In the following, we review several of the main techniques used in the analysis of the coordination networks in this paper.
Although the technical details may vary across different models, the central ideas remain similar.
In the general coordination setting, a network has either classical or quantum communication links. For networks with classical communication links, the main technique is quantum channel resolvability \cite{devetak2003classical,hayashi2016quantum}. For networks with quantum links, we mainly use quantum state redistribution \cite{devetak2008exact,Yard_Devetak_2009,luo2009channel}, Schumacher compression \cite{schumacher1995quantum}, and the quantum version of the classical binning technique~\cite{wyner1975common}. Furthermore, other techniques frequently used in classical information theory are implemented in the quantum setting with modifications, such as random codebook generation and time-sharing \cite{shannon1948mathematical,ElGamalKim:11b}. References to the results presented in this paper are provided with each theorem, where a detailed analysis can be found \cite{natur2025quantum,natur2025empirical,natur2025empirical_ITW}.

\begin{enumerate}[1)]

\item \emph{Channel resolvability:} 
Channel resolvability studies the problem of approximating the output statistics of a channel using a randomly generated codebook \cite{wyner1975common,han1993approximation,hayashi2006general}. 
In the classical setting, the objective is to simulate the output distribution induced by a channel while using as little randomness or communication as possible. 
The quantum extension replaces probability distributions with quantum states and aims to approximate the output state of a classical-quantum (c-q) or a quantum channel \cite{hayashi2016quantum,atif2023quantum}. 
In a two-node network with a noiseless link and limited communication, resolvability determines how much randomness must be pre-shared between users so that, after communication, the joint state of the users is asymptotically indistinguishable from a desired target state.

\item \emph{Quantum state redistribution:} 
In state redistribution, two users initially share parts of a joint quantum state, and the goal is to transfer one subsystem from one user to the other while preserving the global quantum correlations between the subsystems. The analysis characterizes the quantum communication and entanglement resources required to accomplish this transfer.
Quantum state redistribution generalizes several fundamental communication tasks, including state merging and state splitting \cite{devetak2008exact,Yard_Devetak_2009,berta2016smooth}.

The achievability proof for state redistribution by Yard and Devetak \cite{Yard_Devetak_2009} is based on the decoupling theorem, which has become a central tool in quantum Shannon theory. 
The decoupling principle states that quantum information can be transmitted reliably whenever the sender can make the transmitted subsystem nearly independent of the environment. 
Thus, instead of directly reconstructing the desired state at the receiver, it suffices to prove that the environment loses information about the subsystem being transferred. 

\item \emph{Binning:}
Consider a set of binary sequences drawn from a source. In binning, we group subsets of these sequences into indexed groups; each group is called a bin and has an index. The sequences in each bin share a common mathematical property. In communication, rather than sending a complete description of a sequence, the encoder communicates only its bin index, and the decoder uses its correlated side information to identify the correct sequence within that bin. This idea underlies the Slepian--Wolf~\cite{slepian1973noiseless} and Wyner--Ziv~\cite{wyner1976rate} coding schemes and is also used in random binning proofs for coordination and strong secrecy \cite{yassaee2014achievability,cervia2020strong}.

\item \emph{Schumacher compression:} 
Schumacher compression is the quantum analog of classical lossless source coding \cite{schumacher1995quantum,jozsa1994new,barnum1996general}. 
Given many independent realizations of a quantum source, we want to compress the quantum information into the smallest possible Hilbert space while allowing asymptotically reliable reconstruction of the original quantum state. 
The optimal compression rate is characterized by the von Neumann entropy of the source state. 

\end{enumerate}

\section{Information Theoretic Models: Strong Coordination}
\label{Section: Information Theoretic Models: Strong Coordination}
In this section, we study strong coordination in several quantum network topologies.
We first consider networks with quantum communication links, where we study coordination with entangled states. Specifically, we study the broadcast, cascade, and multiple-access networks. 
We then turn to networks with classical communication links and common randomness, where the generated correlations are separable. 
This section illustrates the tradeoff between the different coordination resources and the achievable correlations in each topology.
\subsection{Quantum Links: Entanglement Coordination}
\label{Subsection: Quantum Links: Entanglement Coordination}

\subsubsection{Broadcast Network}
\label{Broadcast_subsection}
\begin{figure}[t]
\center
\includegraphics[scale=0.8,trim={5.3cm 0 5.5cm 0}]
{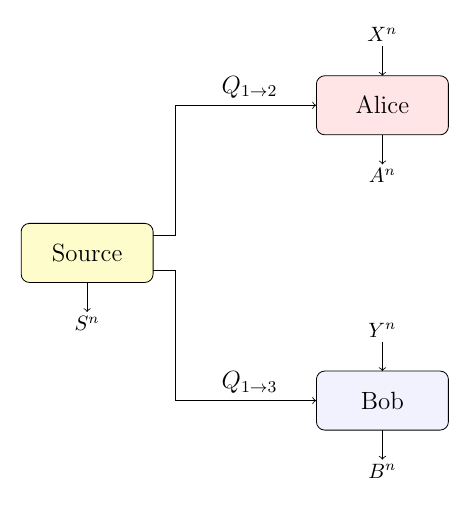} %
\caption{Broadcast network with quantum links.   
The desired joint state is $\omega_{XYSAB}$. Prior to communication, two classical sequences $X^n$ and $Y^n$ are drawn according to a joint probability distribution and provided to Alice and Bob, respectively.
 The Source prepares the state of $S^n$ and the quantum descriptions,  $M_{1 \to 2}$ and $M_{1 \to 3}$, and sends them to Alice and Bob, respectively. Alice uses  $M_{1 \to 2}$ to prepare the state of $A^{n}$, and Bob uses   $M_{1 \to 3}$ to prepare the state of $B^{n}$. 
}
\label{Figure :  Broadcast - Quantum links}
\end{figure}

In the broadcast network presented in Figure~\ref{Figure :  Broadcast - Quantum links}, the Source distributes physical systems to Alice and Bob, who have access to classical side-information. As discussed in Section~\ref{Section: Nonlocal Games}, the model can be applied to determine the resources needed for players to attain a desired winning performance in a refereed game. 
Consider a  classical-quantum (c-q) state,
\begin{align}
\omega_{XYSAB}=\sum_{x\in\mathcal{X}} \sum_{y\in\mathcal{Y}}
p_{XY}(x,y)\ketbra{x,y}_{X,Y}
\otimes \ketbra{\omega^{(x,y)}_{SAB}}
\label{Equation:Broadcast_omega_XYABC}
\end{align}
corresponding to the ensemble $\left\{p_{XY},\ket{\omega^{(x,y)}_{SAB}}\right\}$.  

The users wish to generate a c-q correlation described by the state $\omega_{XYSAB}$.
The protocol begins with $n$ independent realizations of the source distribution $p_{XY}$, forming the sequences $X^n$ and $Y^n$, which are provided to Alice and Bob, respectively.

The Source then prepares the system $S^n$ and the two quantum messages $M_{1\to2}$ and $M_{1\to3}$. Afterwards, he sends the messages to Alice and Bob, respectively, via limited quantum communication links with rates  $Q_{1 \to 2}$ and $Q_{1 \to 3}$.
Alice then uses the message $M_{1 \to 2}$ along with the sequence  $X^n$ to prepare the output  $A^{n}$, and Bob uses the message  $M_{1 \to 3}$ and the sequence $Y^n$ to prepare $B^{n}$. %

\begin{remark}
\label{Remark: Broadcast_quantum_links}
By the no-cloning theorem, it is not possible to broadcast a quantum state to multiple users. However, in the setting shown in Figure~\ref{Figure :  Broadcast - Quantum links}, the source sends different quantum descriptions to Alice and Bob. Basically, the source is broadcasting correlations.
The quantum messages $M_{1 \to 2}$ and $M_{1 \to 3}$ are both prepared by the Source; thus, it is possible to generate entanglement among the three parties.
\end{remark}

\begin{remark}
\label{Remark:Broadcast_no_correlation_quantum_links}
In this network, we assume the Source has no access to the classical sequences; therefore, coordination is possible only for states $\omega_{XYSAB}$ in which $S$ is not correlated with $XY$. i.e., 

\begin{align}
\label{Equation:Broadcast_assumptions}
\omega_{XYS}&=\omega_{XY}\otimes \omega_S \,.
\end{align}
Moreover, because Alice observes only $X^n$ and Bob observes only $Y^n$, the target state must obey the corresponding non-signaling constraints. In particular, Alice's joint state with the source may depend on her own input $x$, but not on Bob's input $y$. Likewise, Bob's joint state with the source may depend on $y$, but not on $x$.

 \end{remark}

The rate requirements for strong coordination in the broadcast network are provided below.

\begin{theorem}[see {\cite[Theorem 5]{natur2025quantum}}]
\normalfont
\label{Theorem: Broadcast - Strong Quantum links}
Consider a state $\omega_{XYSAB}$.  A rate pair $(Q_{1 \to 2},Q_{1 \to 3})$ is achievable for strong coordination in the broadcast network described in Figure~\ref{Figure :  Broadcast - Quantum links}, if and only if
\begin{align}
\begin{array}{rrl}
&Q_{1 \to 2}        &\geq H(A|X)_\omega \,,\\
&Q_{1 \to 3}  &\geq H(B|Y)_\omega\,.
\end{array}
\end{align}
\end{theorem}
The implications of this result on device-independent quantum key distribution (DI-QKD) are discussed in Subsection~\ref{Subsection: Device Independent QKD}.

\begin{remark}
In the broadcast network, the Source, Alice, and Bob can generate any form of entanglement. 
A simple example is a GHZ state, 
$\ket{\omega_{SAB}}=\frac{1}{\sqrt{2}}(\ket{000}+\ket{111})$, taking $X$ and $Y$ to be null (say, $\abs{\mathcal{X}}=\abs{\mathcal{Y}}=1$). In this case, coordination requires $Q_{i \to j}\geq 1$.
\end{remark}

\subsubsection{Cascade Network}
\label{Subsubsection: Cascade Network}

\begin{figure}[t]
\center
\includegraphics[scale=0.8,trim={5.3cm 0 5.5cm 0}]
{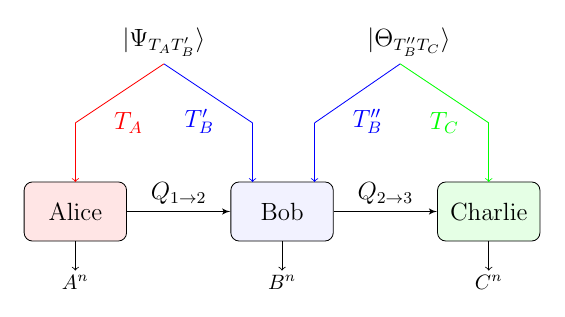} %
\caption{ A quantum-linked Cascade network, with limited communication rates and entanglement assistance. The desired joint state is $\ket{\omega_{RABC}}$.
Prior to communication, entangled qubit pairs are shared between neighboring users.
 Alice prepares the state of $A^{n}$ and the description $M_{1 \to 2}$. She transfers $M_{1 \to 2}$ to Bob. Then Bob uses $M_{1 \to 2}$ to prepare $B^{n}$ and the quantum message $M_{2 \to 3}$, and sends it to Charlie. Charlie then uses it to prepare the output state for $C^{n}$.}
\label{Figure:  Cascade Strong}
\end{figure}

Three users, Alice, Bob, and Charlie, are part of a cascade network as shown in Figure~\ref{Figure:  Cascade Strong}. 
They want to simulate a joint state $\omega_{ABC}$, which is purified by Alice's reference system $R$. The corresponding pure state is $\ket{\omega_{RABC}}$.  
Before communication takes place, neighboring users are provided with bipartite entanglement resources.  
Alice and Bob share the joint state $\ket{\Psi_{T_{A}T_{B}'}}$, while Bob and Charlie share the state $\ket{\Theta_{T_{B}'' T_{C}}}$. The pre-shared entanglement rate between Alice and Bob is $E_{1,2}$, and the rate between Bob and Charlie is $E_{2,3}$.

Initially, Alice prepares the output $A^n$ and the quantum message $M_{1 \to 2}$. She then sends the message to Bob via a quantum link with a limited rate of $Q_{1 \to 2}$. When Bob receives the message, he prepares his output $B^{n}$ and the quantum message $M_{2 \to 3}$. He then sends the message to Charlie, who uses it to prepare the output state of $C^{n}$. 
That is, before the protocol begins, Alice and Bob are provided with $nE_{1,2}$ entangled qubit pairs, while Bob and Charlie share $nE_{2,3}$  pairs. 
During the protocol, Alice transmits $nQ_{1 \to 2}$ qubits to Bob, and then Bob transmits $nQ_{2 \to 3}$ qubits to Charlie. See Figure~\ref{Figure:  Cascade Strong}.

 The optimal rate requirements for coordination over the cascade network are given below. %

\begin{theorem}[see {\cite[Theorem 4]{natur2025quantum}}]
\normalfont
\label{Theorem: Cascade}
Consider a  state $\ket{\omega_{RABC}}$. A rate tuple
$(Q_{1 \to 2},E_{1,2},Q_{2 \to 3},E_{2,3})$ is achievable for
    strong coordination in the cascade network described in Figure~\ref{Figure:  Cascade Strong} if and only if
\begin{align}
\begin{array}{rrl}
& Q_{1 \to 2}       &\geq \frac{1}{2} I(BC;R)_\omega\,,\; 
\\ 
& Q_{1 \to 2}+E_{1,2} &\geq H(BC)_{\omega}\,,
\\ 
& Q_{2 \to 3}  &\geq \frac{1}{2} I(C;RA)_\omega\,,\;
\\ 
&Q_{2 \to 3}+E_{2,3} &\geq H(C)_{\omega}
\,.
\end{array}
\end{align}

\end{theorem}

\begin{corollary}
\normalfont
\label{Theorem_Cascade_Pure}
For a pure state $\ket{\omega_{ABC}}$, 
 a rate tuple
$(Q_{1 \to 2},E_{1,2},Q_{2 \to 3},E_{2,3})$ is achievable in the cascade network under strong coordination if and only if

\begin{align}
\begin{array}{rrl}
& Q_{1 \to 2}+E_{1,2} &\geq H(BC)_{\omega}\,,
\\ 
& Q_{2 \to 3}  &\geq \frac{1}{2} I(C;A)_\omega\,,
\\ 
&Q_{2 \to 3}+E_{2,3} &\geq H(C)_{\omega}
\,.
\end{array}
\end{align}
\end{corollary}

We next illustrate the coordination requirements for two different types of correlation. We focus on the difference between entanglement and separable correlations. 

\begin{figure}[t]
    \centering
    \begin{minipage}[b]{0.4\linewidth} 
        \centering
        \includegraphics[scale=0.58, trim=1cm 0 0 0]{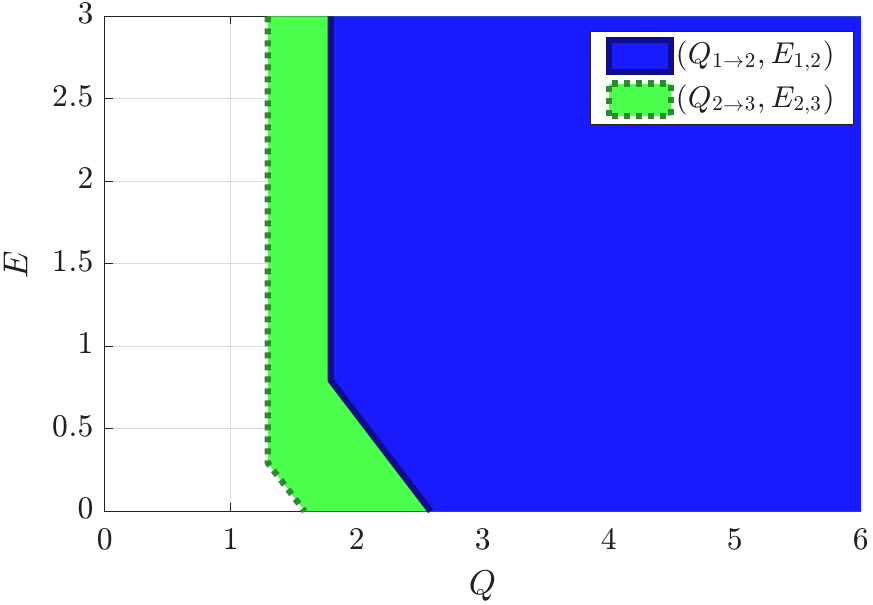} 
        \vspace{-0.5cm}
        \begin{center}
        \footnotesize{(a) Example~\ref{Mixed_example}: Mixture} 
        \end{center}
    \end{minipage}
    \hspace{2.5cm}
    \begin{minipage}[b]{0.4\linewidth} 
        \centering
        \includegraphics[scale=0.58, trim=1.25cm 0 0 0]{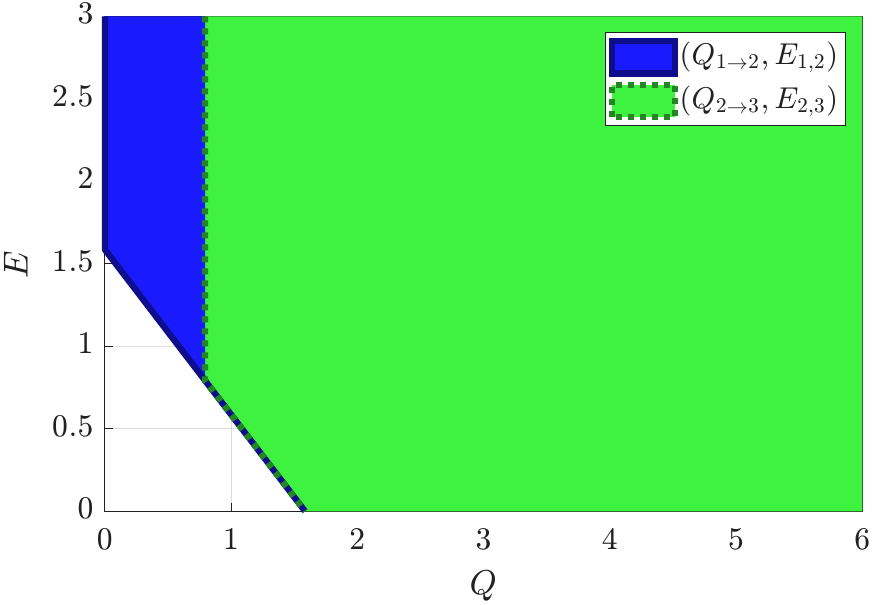} 
        \vspace{-0.5cm}
        \begin{center}
        \footnotesize{(b) Example~\ref{Pure_example}: Entanglement}
        \end{center}
    \end{minipage}
    \caption{Coordination rate regions in two examples. In the first example, subfigure~(a), the state of interest is mixed. The rate of communication from Alice to Bob is \emph{higher} than the rate from Bob to Charlie.   
    In the entangled-state example, subfigure~(b),
    the rate from Alice to Bob is \emph{lower} than the rate from Bob to Charlie.This difference is explained by the “knowing less than nothing” phenomenon, in which the entropy of the full system can be lower than the entropy of its subsystems. 
    }
    \label{Figure_coordination_capacity_cascade}
\end{figure}

\begin{example}
[Mixed state coordination]
\label{Mixed_example}

Consider three qutrit systems, each associated with a three-dimensional Hilbert space. The joint system is described by the following mixed state:
\begin{align}
\hspace{-0.2cm} \omega_{ABC}=\frac{1}{6}\left(\ketbra{012}+\ketbra{021}+\ketbra{102}+\ketbra{120}+\ketbra{201}+\ketbra{210}\right)
\end{align}
A classical analog of this scenario is the task assignment setting \cite[Example 3]{cuff2008communication,cuff2010coordination}.

Based on Theorem~\ref{Theorem: Cascade}, 
the coordination rates required for simulating the mixed state
are illustrated in 
Figure~\ref{Figure_coordination_capacity_cascade} (a). The blue region corresponds to the rate pair $(Q_{1\to2},E_{1,2})$, whereas the green region represents the pair $(Q_{2\to3},E_{2,3})$.
Notice that to establish the desired correlation, Alice 
needs to communicate with Bob at a higher qubit rate than Bob must use to communicate with Charlie.

\end{example}

\begin{example}[Entangled state coordination]
\label{Pure_example}

Consider the following pure tripartite entangled state: 
\begin{align}
\ket{\omega_{ABC}}=\frac{1}{\sqrt{6}}\left(\ket{012}+\ket{021}+\ket{102}+\ket{120}+\ket{201}+\ket{210}\right)
\end{align}
We determine the coordination rates required to generate this correlation using Corollary~\ref{Theorem_Cascade_Pure}.

The rates are illustrated in 
Figure~\ref{Figure_coordination_capacity_cascade} (b). 

In Figure~\ref{Figure_coordination_capacity_cascade} (b),
the rate relation is different from that in the mixed state example.  
In contrast to Example~\ref{Mixed_example}, the required communication rate between Alice and Bob is \emph{lower} than that required from Bob to Charlie.

The examples illustrate that quantum coordination can defy classical intuition.
The difference between classical and quantum coordination arises from entanglement.
To simulate $\omega_{ABC}$, Alice may locally prepare the joint correlation and transmit to Bob the information required to recover \(BC\), after which Bob communicates the subsystem \(C\) to Charlie.
As can be expected, in classical coordination,  Alice must send more information to Bob than Bob sends to Charlie. Indeed, the information sent from Bob to Charlie originates with Alice.
Surprisingly, the opposite occurs in the quantum setting: when the target correlation $\ket{\omega_{ABC}}$ involves entanglement, Alice can send fewer qubits to Bob than Bob sends to Charlie \cite{natur2025quantum}. This counterintuitive behavior arises because entanglement alters the relationship between a system and its subsystems. Although the joint system is in a pure state, its individual subsystems may be described by mixed states and therefore have nonzero entropy.
Furthermore,
while classical conditional entropies are always nonnegative, quantum conditional entropies can become negative in the presence of entanglement.
%
 %
This phenomenon is closely related to the notion of \emph{``knowing less than nothing"} \cite{GourWildeBrandsenGeng:22a}:
A party that already has access to \(A\) learns a negative amount of information when receiving \(BC\), in the sense that $H(BC|A)_\omega<0$.
\end{example}

\subsubsection{Multiple-Access Network}
\label{MAC_subsection}
\begin{figure}[t]
\center
\includegraphics[scale=0.8,trim={5.3cm 0 5.5cm 0}]
{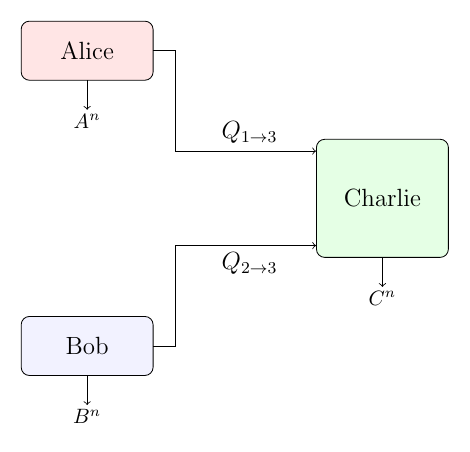} %
\caption{Multiple-access network with quantum links.  The users aim to simulate a desired quantum state $\omega_{ABC}$. Alice prepares the output state of  $A^{n}$, and the quantum message  $M_{1 \to 3}$. 
Likewise, Bob prepares the state of $B^{n}$ and $M_{2 \to 3}$.
They send $M_{1 \to 3}$ and $M_{2 \to 3}$ to Charlie. Upon receiving the quantum messages, Charlie prepares the output $C^n$. %
}
\label{Figure :  Multiple-access}
\end{figure}

Alice, Bob, and Charlie are users in the multiple-access network presented in Figure~\ref{Figure :  Multiple-access}. They wish to simulate a quantum correlation described by the state %
$\ket{\omega_{ABC}}^{\otimes n}$.
The protocol proceeds as follows.
Alice starts by preparing the states of her systems $A^{n}$ and the quantum description $M_{1\to3}$. Similarly, Bob prepares the quantum state of $B^{n}$ and creates his quantum description $M_{2\to3}$. They both send their corresponding descriptions, $M_{1\to3}$ and $M_{2\to3}$, to Charlie. Upon receiving the descriptions, Charlie prepares the state of $C^{n}$. The communication links are limited to qubit rates of $Q_{1\to3}$ and $Q_{2\to3}$ from Alice to Charlie and Bob to Charlie, respectively.

\begin{remark}
\label{MAC_product_state}
Since Alice and Bob prepare their messages independently, and they do not have pre-shared resources, the systems they send cannot be correlated. Therefore, the protocol can generate only target states satisfying
\begin{align}
\label{Equation: MAC_product}
\omega_{AB}=\omega_A\otimes\omega_B.
\end{align}
In particular, if the target state contains correlations between Alice and Bob, then it cannot be generated over the multiple-access network.
\end{remark}

The optimal rate requirements for coordination over the multiple-access network are given below.

\begin{theorem}[see {\cite[Theorem 6]{natur2025quantum}}]
\normalfont
\label{Theorem: Multiple-access}
Let $\ket{\omega_{ABC}}$ be a pure state satisfying the condition in \eqref{Equation: MAC_product}.
A rate pair $(Q_{1 \to 3},Q_{2 \to 3})$ is achievable for strong coordination in the multiple-access network described in Figure~\ref{Figure :  Multiple-access} if and only if
\begin{align}
\begin{array}{rrl}
&
Q_{1 \to 3}        &\geq H(A)_\omega \,,\\
&Q_{2 \to 3}  &\geq H(B)_\omega\,.
\end{array}
\end{align}
\end{theorem}
\begin{remark}
The following simple cases provide further insight into the rate requirements.
For the state $\ket{\omega_A}\otimes \ket{\omega_{BC}}$,  the rate from Alice to Charlie is $Q_{1 \to 3}=0$, since no communication is needed between Alice and Charlie to achieve coordination. 
For $\ket{\omega_{AC}}\otimes \ket{\omega_{B}}$, the rate from Bob to Charlie is $Q_{2 \to 3}=0$.
Another interesting case is the simulation of a maximally entangled state $\omega_{AC}$ between Alice and Charlie. In this case, the required coordination rates are $Q_{1\to3}\geq1$ and $Q_{2\to3}\geq0$.
Additionally, if $\ket{\omega_{ABC}}=\ket{\omega_{AB}}\otimes \ket{\omega_{C}}$, then $\omega_{AB}=\ketbra{\omega_A}\otimes\ketbra{\omega_B}$. Therefore, no communication is needed to establish coordination.
\end{remark}

\begin{remark} 
\label{Remark: repeaters}
Assume that Alice, Bob, and Charlie want to simulate two pairs of maximally entangled qubits $\ket{\omega_{ABC_1 C_2}}=\ket{\phi_{AC_1}}\otimes \ket{\chi_{BC_2}}$, where $C\equiv 
({C_1},{C_2})
$. One pair is between Alice and Charlie and the other is between Bob and Charlie. According to the coordination result in the multiple-access network in Theorem~\ref{Theorem: Multiple-access}, the communication rates must satisfy $Q_{i \to j}\geq 1$.
Once the two pairs have been generated, Charlie performs a local Bell measurement on systems $C_1$ and $C_2$. Entanglement swapping then produces a maximally entangled pair between Alice and Bob.
Further discussion about the importance of coordination in repeater networks is provided in Subsection~\ref{Subsection: Quantum Repeaters}.
\end{remark}

\subsection{Classical Links: Separable Correlations}
\label{Subsection: Classical Links: Separable Correlations}

\subsubsection{Broadcast Network}
\label{Subsection:BC}
\begin{figure}[t]
\center
\includegraphics[scale=0.85,trim={5.8cm 0 5.5cm 0}]
{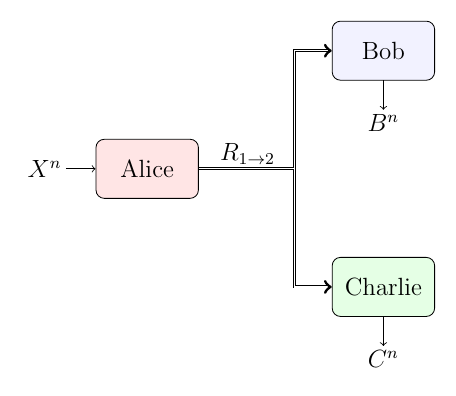} 
\caption{Broadcast Network with classical links and common randomness assistance. 
Alice, Bob, and Charlie are provided with a common randomness element  $m_0$ before the protocol begins.
A classical sequence
$X^n$ is generated by Nature and given to Alice. 
She then sends a message $m_{1\to 2}$ to Bob and Charlie.
As Bob and Charlie receive the message, they encode the state of their respective output, $B^n$ and $C^n$.
The desired c-q state is $\omega_{XBC}$.
The CR element is omitted from the figure for simplicity. 
}
\label{Figure : Broadcast - Classcial links}
\end{figure}

In the broadcast network shown in Figure~\ref{Figure : Broadcast - Classcial links}, the sender Alice and the receivers Bob and Charlie aim to simulate the c-q state $\omega_{XBC}$. They follow the protocol below. The users share a CR element $m_0$ at a rate $R_{0}$. A classical source described by a PMF $p_{X}$ provides Alice with an i.i.d. sequence $x^{n}\in\mathcal{X}^{n}$. The two other users, Bob and Charlie, have quantum systems $B$ and $C$. Alice uses the sequence $x^n$ to generate a classical message $m_{1\to2}$ at a rate $R_{1\to2}$ and sends the same message to both Bob and Charlie.

Consider a c-q state $\omega_{XBC}$, and denote the  following
 set of c-q extensions by $\mathscr{S}_{\text{Broadcast}}(\omega)$ 
    \begin{align}
    \sigma_{XUBC}	&=\sum\limits _{\substack{(x,u)\in \\ \mathcal{X}\times\mathcal{U}}}p_{XU}(x,u)\ketbra{x}_{X}
  \otimes\ketbra{u}_{U}\otimes\theta_{B}^{u}\otimes\eta_{C}^{u}
\end{align}
such that 
\begin{align}
 \sigma_{XBC}&=\omega_{XBC} \,.
\end{align}
 Notice that given $U=u$,  
there is no correlation between $X$, $B$, and $C$.  
\begin{theorem}\cite[Theorem 3]{natur2025quantum}
\label{Theorem: Broadcast - Classcial links}
Consider a state $\omega_{XBC}$. A rate pair $(R_0,R_{1\to2})$
is achievable for strong coordination in the broadcast network in Figure~\ref{Figure : Broadcast - Classcial links}, if and only if
\begin{align}
\begin{array}{rrl}
&
R_{1\to2}        &\geq I(X;U)_\sigma \,,\\
&R_0+R_{1\to2}  &\geq I(XBC;U)_\sigma\,.
\end{array}
\end{align}
\end{theorem}


\begin{remark}
\label{Remark:Entanglement_BC}
Alice's encoding is classical; therefore, entanglement cannot be distributed. Furthermore, the absence of cooperation between Bob and Charlie prevents them from generating entangled correlations.
 Thus $\omega_{BC}$ must be separable, as in the no-communication model (see Remark~\ref{Remark:Entanglement_2}). 
\end{remark}

\subsection{No-Communication Network}
\label{Subsection: No-Communication Network}
 
\begin{figure}[t]
\center
\includegraphics[scale=0.8,trim={3.75cm 0 4cm 0}]
{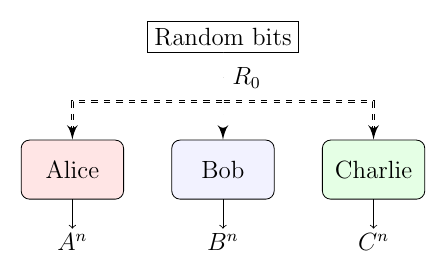} 
\caption{
No-communication network. %
%
Alice, Bob, and Charlie are only provided with a common randomness element  $m_0$, and cannot communicate with one another. 
Each encoder applies a local encoding map to encode their outputs $A^n$, $B^n$, and $C^n$, respectively.
The goal is to asymptotically simulate a joint quantum state $\omega_{ABC}$.}
\label{Figure : No-communication}
\end{figure}

Alice, Bob, and Charlie hold systems $A$, $B$, and $C$, respectively. They are unable to communicate; however, they have access to pre-shared randomness at a rate $R_0$. The CR element is denoted by $m_0$ as depicted in Figure ~\ref{Figure : No-communication}. Using the common-randomness element, each user prepares the corresponding system independently.

The optimal CR rate for strong coordination in the no-communication network is presented below.

Consider a quantum state $\omega_{ABC}$. 
Introduce the set $\mathscr{S}_{\text{NC}}(\omega)$ of all c-q extensions of the following form. 
\begin{subequations}
 \label{Equation:S_NC}
\begin{align}
  \sigma_{UABC}&=
 \sum\limits_{u\in\mathcal{U}}p_{U}(u)
 \; 
 \ketbra{u}_{U}\otimes \theta_{A}^{u}\otimes\theta_{B}^{u}\otimes\theta_{C}^{u} 
\end{align}
such that 
\begin{align}
 \sigma_{ABC}&=\omega_{ABC}
 \,.
\end{align}
\end{subequations}
Notice that systems $A$,$B$, and $C$ are uncorrelated given $U=u$.

\begin{theorem}\cite[Theorem 2]{natur2025quantum}
\label{Theorem: No-communication}
A rate $R_0$ is achievable for strong coordination in the no-communication network in Figure~\ref{Figure : No-communication} if and only if 
\begin{align}
R_0\geq
\inf_{\sigma_{UABC}\in\mathscr{S}_{\text{NC}}(\omega)}
 I(U;ABC)_\sigma
 \label{Equation:No_Communication}
\end{align}
with the convention that an infimum over an empty set  is $+\infty$.
\end{theorem} 

 The no-communication network was studied independently by George et al. \cite[Sec. VII]{george2023one}.
\begin{remark}
\label{Remark:Entanglement_2}
In the absence of pre-shared entanglement or quantum communication, common randomness alone cannot be used to generate entanglement. Given that the users do not cooperate during the coordination protocol, they cannot generate entangled correlations, and the simulation is limited to separable states.
\end{remark}

\begin{remark}
For a product state $\omega_{ABC}=\omega_A\otimes \omega_B \otimes \omega_C$, we may take $U$ to be null, hence the condition for coordination is $R_0\geq 0$. That is, 
simulation does not require CR between the users. 
On the other hand, if $\omega_{AB}$ is entangled,  then there is no $U$ that can satisfy \eqref{Equation:S_NC}, and coordination is not possible. 
    For a classically correlated state, 
 $
        \omega_{ABC}=\frac{1}{2}\left(\ketbra{000}+\ketbra{111}\right)
  $, 
we need $R_0\geq1$, as one bit of CR is required in order to simulate such a correlation. 
\end{remark}

\section{Information Theoretic Models: Empirical Coordination}
\label{Section: Information Theoretic Models: Empirical Coordination}

This section reviews information-theoretic models for empirical coordination in quantum networks. 
In contrast to strong coordination, where the objective is to reproduce the full behavior of a memoryless source, empirical coordination focuses only on the average statistical behavior of the generated systems. 
We consider both quantum and classical communication links and discuss how the available resources determine the set of achievable correlations. 
The examples presented in this section illustrate two complementary settings. 
First, we study entanglement coordination over quantum communication links, where the users generate quantum correlations assisted by side information. 
We then consider separable correlations over classical communication links, highlighting the differences between classical and quantum coordination mechanisms. 
Together, these models demonstrate how empirical coordination provides a unified framework for characterizing the communication resources required for generating distributed quantum correlations.

\subsection{Quantum Links: Entanglement Coordination}
\label{Quantum Links: Empirical Entanglement Coordination}

Consider a c-q state of the following form, %
\begin{align}
\omega_{XYSAB}=\sum_{x\in\mathcal{X}} \sum_{y\in\mathcal{Y}}
p_{XY}(x,y)\ketbra{x,y}_{X,Y}
\otimes \ketbra{\omega^{(x,y)}_{SAB}}\,,
\label{Equation:Empirical_Broadcast_omega_XYABC}
\end{align}
where $X$ and $Y$ are random variables with a joint probability distribution $p_{XY}$.
In the quantum-linked broadcast network shown in Figure~\ref{Figure :  Broadcast - Quantum links}, 
the Source, Alice, and Bob aim to generate a correlation described by a joint state $\omega_{XYSAB}$ on average.

Prior to communication, the classical sequences $X^n$ and $Y^n$ are jointly generated according to the product distribution $p_{XY}^{\otimes n}$. Alice is provided with $X^n$, while Bob is provided with $Y^n$, as depicted in Figure~\ref{Figure : Broadcast - Quantum links}. 
To establish empirical coordination, they use the following scheme.  
The Source first prepares the systems $S^{n}$, together with two quantum systems, $M_{1\to 2}$ and $M_{1 \to 3}$. $M_{1 \to 2}$ is sent to Alice with a qubit transmission rate of $Q_{1\to 2}$, while $M_{1 \to 3}$ is sent to Bob at a rate of  $Q_{1\to 3}$.
Using the quantum description $M_{1\to 2}$ and the sequence $X^n$, Alice prepares the systems $A^{n}$. Likewise, Bob uses $M_{1\to 3}$ together with $Y^n$ to prepare the systems $B^n$.

The observations in Remark~\ref{Remark: Broadcast_quantum_links} for strong coordination over the quantum broadcast network also apply to the empirical setting, since they follow from the network topology.

The rate requirements for empirical coordination in the broadcast network with quantum links are established below.
\begin{theorem}\cite[Theorem 4]{natur2025empirical_ITW}
\normalfont
\label{Theorem: Broadcast Empirical}
Let $\omega_{XYSAB}$ be as in \eqref{Equation:Empirical_Broadcast_omega_XYABC}, such that \eqref{Equation:Broadcast_assumptions} holds.
 A rate pair $(Q_{1 \to 2},Q_{1 \to 3})$ is achievable for empirical coordination in the broadcast network described in Figure~\ref{Figure :  Broadcast - Quantum links}, if and only if
\begin{align}
\begin{array}{rrl}
&
Q_{1\to 2}        &\geq H(A|X)_\omega \,,\\
&Q_{1\to 3}  &\geq H(B|Y)_\omega\,.
\end{array}
\end{align}
\end{theorem}

\subsection{Classical Links: Separable Correlations}
\label{Subsection: Classical Links: Empirical Separable Correlations}
\begin{figure}[tb]
\center
\includegraphics[scale=0.75,trim={5.3cm 0 5.5cm 0}]
{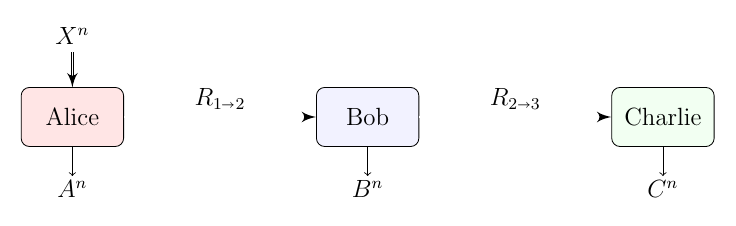} %
\caption{
A classically linked cascade network with no pre-shared randomness. Alice observes the output sequence \(X^n\) of a classical source characterized by the probability distribution \(p_X\), prepares her system \(A^n\), and sends a classical message \(m_{1\to2}\) to Bob at rate \(R_{1\to2}\) through a classical communication link. Upon receiving \(m_{1\to2}\), Bob prepares his system \(B^n\) and sends a classical message \(m_{2\to3}\) to Charlie at rate \(R_{2\to3}\). Upon receiving \(m_{2\to3}\), Charlie prepares his system \(C^n\).
}
\label{Figure: Cascade network}
\end{figure}
%

In the cascade network presented in Figure~\ref{Figure: Cascade network}, 
Alice, Bob, and Charlie aim to generate a separable correlation described by the joint state $\omega_{ABC}$. The protocol proceeds as follows.
A memoryless source $p_X$ provides Alice with a sequence of classical side information. Upon receiving it, she %
prepares the state of $A^n$, chooses 
 an index $m_{1 \to 2}$ and sends it to Bob; this is her classical message. The communication rate is limited to $R_{1 \to 2}$. Once Bob receives the message $m_{1 \to 2}$, he encodes the systems $B^n$ and then sends $m_{2 \to 3}$ to Charlie, enabling him to prepare the systems $C^n$.

The rate requirements for empirical coordination in the cascade network with classical links are given below. Consider the extended c-q state,
\begin{align}
\omega_{XABC}=
\sum_{x\in\mathcal{X}}
p_X(x) \ketbra{x}_X\otimes \omega_{ABC}^x \,.
\end{align}
Furthermore, denote by $\mathscr{S}_{\text{Cascade}}(\omega)$ the set of all c-q extensions of the form  
\begin{subequations}
\begin{align}
  \sigma_{XYZABC}&=
 \sum\limits_{
 (x,y,z)\in\mathcal{X}\times \mathcal{Y}\times \mathcal{Z}
 }
 p_{XYZ}(x,y,z)  
 \ketbra{x}\otimes\ketbra{y}\otimes\ketbra{z}\otimes \sigma_A^x
  \otimes 
  \sigma_{B}^{y}
  \otimes 
  \sigma_{C}^{z}
\end{align}
such that %
\begin{align}
 \sigma_{XABC}&=\omega_{XABC} \,.
\end{align}
\end{subequations}

Notice that coordination over classical links is restricted to separable correlations, since generating entanglement requires pre-shared entanglement resources or quantum links.

The rate requirements for empirical coordination in the cascade network with classical links are provided below.
\begin{theorem}\cite[Theorem 3]{natur2025empirical}
\label{Theorem: Empirical Cascade }
Consider a three-party state $\omega_{ABC}$.
    If the set $\mathscr{S}_{\text{Cascade}}(\omega)$ is nonempty, then a rate pair $(R_{1 \to 2},R_{2 \to 3})$ is achievable for empirical coordination in the cascade network  in Figure~\ref{Figure: Cascade network}, if and only if 
\begin{align}
\begin{array}{rrl}
&
R_{1 \to 2}        &\geq I(X;YZ)_\sigma \,,\\
&R_{2 \to 3}  &\geq I(X;Z)_\sigma\,.
\end{array}
\end{align}
Otherwise, if $\mathscr{S}_{\text{Cascade}}(\omega) = \emptyset$, then coordination is impossible.
\end{theorem}

\begin{remark}
Notice that once Charlie has Bob's message, $m_{2\to 3}$, Charlie's state has no correlation with Alice. This gives the communication structure of the cascade network a Markov form. However, the produced correlation between the users does not follow a Markov chain structure. In particular, although the joint distribution of $X$, $Y$, and $Z$ obeys the usual probability rules, these random variables need not form a Markov chain. 
\end{remark}

\section{Discussion and Conclusions}
\label{Section: Discussion and Conclusions}

In this section, we consider additional applications of quantum coordination.
We begin with device-independent quantum key distribution (DI-QKD), where the generation of nonlocal correlations is a resource for secure communication. We then discuss quantum repeaters, focusing on the role of coordination in generating and distributing long-range entanglement. 
Finally, we conclude the review and outline future directions, including large-scale quantum networks, finite blocklength coordination, multipartite entanglement generation, and additional applications in distributed quantum systems.

\subsection{Device-Independent Quantum Key Distribution}
\label{Subsection: Device Independent QKD}
The main objective in coordination is to generate nontrivial correlations among spatially separated users using limited communication and shared resources. 
In the quantum setting, such correlations may arise from entanglement and exhibit nonlocal behavior that has no classical counterpart. 
These correlations are a valuable resource for cryptographic and communication tasks, especially in device-independent quantum key distribution (DI-QKD).

In DI-QKD, security does not rely on assumptions regarding the internal implementation of the devices used by the legitimate parties, but rather on the generation of nonlocal correlations between them. These correlations are typically certified by observing a violation of a Bell inequality. Thus, the ability to generate desired correlations is essential for quantum cryptography.

We begin our discussion of DI-QKD with the protocol introduced in \cite{acin2007device,pironio2009device}, which extends the original entanglement-based quantum key distribution scheme proposed by Ekert \cite{ekert1991quantum}.
The protocol operates as follows. It is carried out over $N$ rounds, which are divided into
key-generation and test rounds.
In each round, a potentially malicious source, assumed to be under Eve’s control, distributes quantum systems to Alice and Bob, who choose local measurements and record their outcomes.
The key-generation rounds provide the raw data
used to construct the secret key, whereas the test rounds are used to
estimate the CHSH violation and assess the behavior of the devices.
Alice and Bob publicly reveal part of their data to estimate the
relevant security parameters. If the observed values satisfy the
required thresholds, they apply error correction and privacy
amplification to obtain a shared secret key.

Alice has access to three measurement settings, denoted by $A_0$, $A_1$, and $A_2$, each producing outcomes in $\{-1,+1\}$. 
Bob performs one of two possible measurements, $B_0$ or $B_1$, also with outputs in $\{-1,+1\}$. 
The raw key is generated from the outcomes associated with the measurement pair $(A_2,B_0)$. 
These outcomes are then used for classical post-processing procedures, such as error correction and privacy amplification. 
The statistics obtained from this measurement pair determine the quantum bit error rate (QBER), which is the probability that Alice and Bob get different measurement results when measuring $(A_2,B_0)$. We denote the error rate by $q$, namely,
\begin{align}
    q=P_{AB|XY}(a\neq b|2,0)\,.
\end{align} 
In the test rounds, the measurements $A_0$, $A_1$, $B_0$, and $B_1$
are used to evaluate the CHSH expression, namely,
\begin{align}
T=\langle A_0 B_0\rangle+\langle A_0 B_1\rangle+\langle A_1 B_0\rangle-\langle A_1 B_1\rangle,
\end{align}
where $\langle A_x B_y\rangle$ denotes the correlation between the outcomes corresponding to the measurement settings indexed by $x$ and $y$.
The value of $T$ quantifies the observed Bell violation. 
Together, the parameters $q$ and $T$ allow the legitimate users to estimate the information potentially available to Eve.
We treat parameter estimation as a separate stage to emphasize the central role of Bell-inequality verification in DI-QKD.

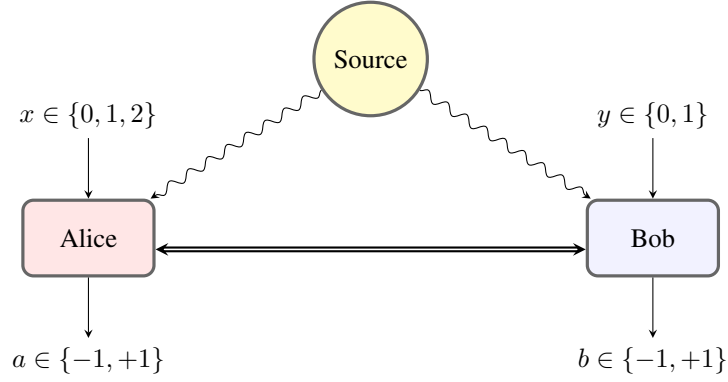
\begin{figure}[t]
    \centering
    \begin{tikzpicture}[
        node distance=2.2cm,
        aliceblock/.style={
            rectangle,
            draw=black!60,
            very thick,
            fill=red!10,
            text width=1.5cm,
            text centered,
            rounded corners,
            minimum height=1cm
        },
        bobblock/.style={
            rectangle,
            draw=black!60,
            very thick,
            fill=blue!5,
            text width=1.5cm,
            text centered,
            rounded corners,
            minimum height=1cm
        },
        sourcenode/.style={
            circle,
            draw=black!60,
            very thick,
            fill=yellow!25,
            minimum size=1.5cm
        },
        >=stealth
    ]

    \node[sourcenode] (source) {Source};

    \node[aliceblock]
        (alice)
        [below left=1.3cm and 2.3cm of source]
        {Alice};

    \node[bobblock]
        (bob)
        [below right=1.3cm and 2.3cm of source]
        {Bob};

    \node (x)
        [above=0.8cm of alice]
        {$x\in\{0,1,2\}$};

    \node (a)
        [below=0.8cm of alice]
        {$a\in\{-1,+1\}$};

    \node (y)
        [above=0.8cm of bob]
        {$y\in\{0,1\}$};

    \node (b)
        [below=0.8cm of bob]
        {$b\in\{-1,+1\}$};

    \draw[
        ->,
        decorate,
        decoration={
            snake,
            amplitude=0.6mm,
            segment length=3mm
        }
    ]
    (source) -- (alice);

    \draw[
        ->,
        decorate,
        decoration={
            snake,
            amplitude=0.6mm,
            segment length=3mm
        }
    ]
    (source) -- (bob);

    \draw[->] (x) -- (alice);
    \draw[->] (alice) -- (a);

    \draw[->] (y) -- (bob);
    \draw[->] (bob) -- (b);

    \draw[<->,double ,thick]
        ([yshift=-0.15cm]alice.east)
        --
        ([yshift=-0.15cm]bob.west);

    \end{tikzpicture}

    \caption{
    DI-QKD protocol.
    A source distributes quantum systems to Alice and Bob.
    Alice chooses a measurement $x\in{0,1,2}$ and obtains an outcome $a\in{-1,+1}$, while Bob chooses $y\in{0,1}$ and obtains an outcome $b\in{-1,+1}$.
    Alice and Bob also have access to an authenticated classical communication channel $C$.
    }    \label{Figure:DIQKD_protocol}
\end{figure}

\paragraph*{DI-QKD Protocol}

The protocol is parameterized by the total number of rounds $N$, a testing fraction $\gamma\in[0,1]$, and a noise tolerance parameter $\eta$.

\begin{enumerate}
    \item \emph{Measurement stage:}
    
    For each round $i\in\{1,\ldots,N\}$, Alice and Bob independently choose a measurement 
    \begin{align}
        x_i\in\{0,1,2\},
        \qquad
        y_i\in\{0,1\},
    \end{align}
    uniformly at random. They apply the corresponding measurements to their devices and obtain outcomes
    \begin{align}
        a_i,b_i\in\{+1,-1\}.
    \end{align}

   \textit{2) Parameter-estimation stage:}

After completing the $N$ rounds, Alice randomly chooses a test set
\begin{align}
\mathfrak{T} \subseteq [N]\,,
\qquad
|\mathfrak{T}| = \gamma N\,.
\end{align}
She publicly announces the indices in
$\mathfrak{T}$.

For every round $i\in\mathfrak{T}$, Alice and Bob reveal their chosen
measurements $(x_i,y_i)$ and their corresponding outcomes $(a_i,b_i)$.
These rounds are used only for parameter estimation and are
discarded afterward.

Using the revealed data, Alice and Bob estimate the CHSH parameter
$T$. They abort the protocol if the estimated Bell violation is below a specific threshold. Furthermore, they estimate the disagreement probability for the
key-generation measurements $(x,y)=(2,0)$, and abort if it exceeds the
tolerated error rate $\eta$.

\textit{3) Key-extraction stage:}

If Alice and Bob did not abort during the parameter-estimation stage, they retain the remaining rounds in which the key-generation measurements were performed.
Define
\begin{align}
\mathfrak{S}
:=
\left\{
i\in[N]\setminus\mathfrak{T}
:
(x_i,y_i)=(2,0)
\right\}.
\end{align}

Alice's outcomes $\{a_i\}_{i\in\mathfrak{S}}$ and Bob's outcomes
$\{b_i\}_{i\in\mathfrak{S}}$ form their raw keys. They then perform classical
information reconciliation to correct mismatches between the raw keys,
followed by privacy amplification to remove information that may be available
to an adversary. The result is a shared secret key.
\end{enumerate}

\subsection{Example: DI-QKD under Depolarizing Noise }

To illustrate the device-independent protocol, we consider the following implementation, following the setup in~\cite{ekert1991quantum}.
Suppose that the Source initially prepares the  following pure 
state
\begin{align}
\ket{\omega}_{AB}
=
\sqrt{\alpha}\ket{00}
+
\sqrt{1-\alpha}\ket{11},
\qquad
0\leq\alpha\leq 1.
\label{Equation:initial_nonmax_state}
\end{align}
The distributed state is subjected to global depolarizing noise,
described by
\begin{align}
\mathcal{D}_{p}(\tau_{AB})
=
(1-p)\tau_{AB}
+
p\frac{\identity_{AB}}{4}\operatorname{Tr}(\tau_{AB}),
\qquad
0\leq p\leq 1,
\end{align}
where $p$ denotes the depolarization probability. Therefore, the state shared by
Alice and Bob is 
\begin{align}
\rho_{AB}^{(\alpha,p)}
=
(1-p)\ket{\omega_{\alpha}}\bra{\omega_{\alpha}}
+
p\frac{\identity_{AB}}{4}.
\label{Equation:depolarized_nonmax_state}
\end{align}

Alice and Bob apply the measurements
\begin{align}
\begin{aligned}
A_0&=\frac{1}{\sqrt{2}}(Z+X),
&\qquad
B_0&=Z,
\\[0.3em]
A_1&=\frac{1}{\sqrt{2}}(Z-X),
&
B_1&=X,
\\[0.8em]
A_2&=Z.
\end{aligned}
\end{align}

For the state in \eqref{Equation:depolarized_nonmax_state}, the CHSH parameter is
\begin{align}
T(\alpha,p)=
\sqrt{2}(1-p)
\left(
1+2\sqrt{\alpha(1-\alpha)}
\right).
\label{Equation:CHSH_alpha_p}
\end{align}

The QBER is
\begin{align}
q
=
\frac{p}{2}.
\label{Equation:QBER_alpha_p}
\end{align}

As in standard QKD security analyses, Eve is assumed to possess a quantum system that may be correlated with those held by Alice and Bob. The asymptotic
secret-key rate is lower bounded by the Devetak--Winter expression
\cite{devetak2005distillation},
\begin{align}
R_{\mathrm{key}}
\geq
I(A_2;B_0)-\chi(B_0;E),
\end{align}
where $I(A_2;B_0)$ quantifies the correlation between the
key-generation outcomes and $\chi(B_0;E)$ bounds Eve's accessible
information.

For uniform marginals, the corresponding device-independent lower bound
is \cite{wolf2021quantum}
\begin{align}
R_{\mathrm{key}}
\geq
1-h_2(q)
-
h_2\left(
\frac{
1+\sqrt{\left(\frac{T}{2}\right)^2-1}
}{2}
\right),
\label{Equation:general_DI_key_rate_example2}
\end{align}
where $h_2(\cdot)$ denotes the binary entropy function.

Based on Theorem~\ref{Theorem: Broadcast - Strong Quantum links}, 
the
quantum communication rates required to generate
$\rho_{AB}^{(\alpha,p)}$ must satisfy

\begin{align}
Q_{1\to 2}\,,\;Q_{1\to 3}
\geq
h_2\left(
(1-p)\alpha+\frac{p}{2}
\right),
\label{Equation:DIQKD_coordination_rates}
\end{align}

\begin{figure}[t]
    \centering
    \includegraphics[width=0.75\linewidth]{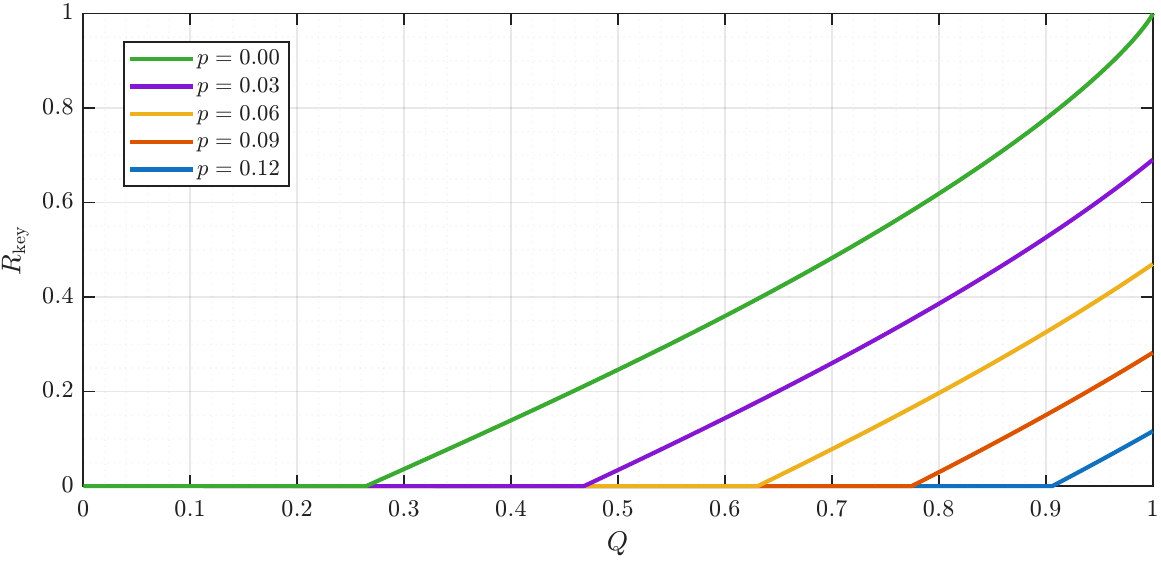}
    \caption{
    Device-independent secret-key rate $R_{\mathrm{key}}$ as a function
    of the quantum communication rate
    $Q_{1\to 2}=Q_{1\to 3}\equiv Q$ for different values of the depolarization
    probability $p$.
    }
    \label{Figure: DI_key_rate}
\end{figure}

For $p=0$, the state is pure, and the rates reduce to
\begin{align}
Q_{1\to 2}\,,\;Q_{1\to 3}
\geq
h_2(\alpha)
\end{align}
in agreement with the pure-state case considered in
Subsection~\ref{Subsection: CHSH}. In Figure~\ref{Figure: DI_key_rate}, we plot the device-independent key rate as a function of the coordination rate $ Q_{1\to2}= Q_{1\to3}\equiv Q$, for $\alpha\in[0,\frac{1}{2}]$ and several values of $p$.
Notice that for $p=0$, to achieve a Bell violation when $\alpha=\frac{1}{2}$, the source needs to send qubits at rates $Q_{1\to2},Q_{1\to3}\geq 0.2643$. In Figure~\ref{Figure: DI_key_rate}, this is the point at which the curve corresponding to $p=0$ becomes positive. This result is consistent with the discussion of the CHSH game in Subsection~\ref{Subsection: CHSH}.

\subsection{Quantum Repeaters}
\label{Subsection: Quantum Repeaters}

Communicating quantum information depends on the long-distance transmission of quantum systems \cite{munro2015inside}. Factors such as signal attenuation and loss severely limit long-range communication  \cite{pereg2023communication}. 
Classical communication systems can compensate for such losses by amplifying the signal. However, this approach does not apply to quantum communication, as the no-cloning theorem prohibits the replication of an unknown quantum state~\cite{wootters1982single}. 
Therefore, alternative methods are required to extend the range of quantum communication networks.

The central idea behind a quantum repeater is to avoid transmitting a fragile quantum state directly over the entire communication distance. 
Instead, the total distance is divided into shorter elementary links. 
Entanglement is first generated between neighboring nodes and stored locally. 
Intermediate nodes then perform Bell measurements that swap the entanglement from neighboring links into a longer-distance entangled state. 
By repeatedly applying this procedure, two distant users can eventually share entanglement even though no quantum system has traveled the entire communication path. 
The resulting long-range entanglement can subsequently be used for quantum teleportation, entanglement-based quantum key distribution, and other distributed quantum communication tasks \cite{bennett1993teleporting,briegel1998quantum}.

The need for such a procedure can be illustrated through a simple numerical example. 
Modern telecommunication fibers typically have an attenuation of approximately $0.2$ dB/km at a wavelength of approximately $1.5,\mu\mathrm{m}$ \cite{sangouard2011quantum}. 
Although this corresponds to a transmission probability of roughly $95\%$ over a distance of $1$ km, the probability of successful transmission decreases exponentially with distance. 
Even for a source repetition rate of $10$ GHz, direct transmission over $500$ km would yield only about one successfully transmitted photon per second. 
At $600$ km, the rate drops to approximately $0.01$ Hz, and at $1000$ km it becomes roughly $10^{-10}$ Hz, corresponding to about one photon every few hundred years \cite{sangouard2011quantum}. 
Quantum repeaters overcome this severe distance-dependent loss by generating entanglement over shorter links and extending it through entanglement swapping and purification.

\begin{figure}[t]
\center
\includegraphics[scale=0.75,trim={5.3cm 0 5.5cm 0}]
{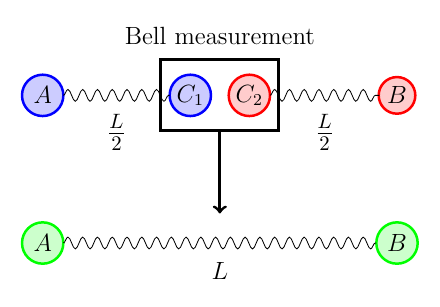} 
\caption{A basic quantum repeater network \cite{van2020extending}.
The protocol begins with systems $C_1$ and $C_2$ in Charlie's possession. Alice's system $A$ and system $C_1$ are maximally entangled, and Bob's system $B$ and system $C_2$ are maximally entangled as well. In each pair, the distance between systems is $L/2$. A local Bell measurement on Charlie’s systems produces entanglement swapping, leaving $A$ and $B$ maximally entangled across the full distance $L$.}
\label{Figure: Repeaters}
\end{figure}

This severe distance-dependent loss motivates the use of quantum repeaters, which replace direct transmission with a sequence of shorter elementary links \cite{briegel1998quantum}. 
The protocol is initiated by the creation of Bell pairs between adjacent nodes. The entanglement is then extended over longer distances through entanglement swapping. For illustration, see Figure~\ref{Figure: Repeaters}.
The long-distance Bell pair enables the transmission of quantum information via quantum teleportation from the sender to the receiver. Considering their significant potential, the development of large-scale quantum networks is expected to depend heavily on repeater-assisted transmission over long distances \cite{goodenough2021optimizing}. Accordingly, quantum repeaters have been investigated using several experimental platforms and physical implementations~\cite{wilksen2024gate,neuwirth2021quantum,bergerhoff2024quantum,dhara2022multiplexed,mizuochi2024quantum,azuma2023quantum,rodgers2021materials}.
Given the reliance of quantum repeaters on nontrivial entanglement correlations, quantum coordination is central for their implementation. 
For instance, consider the multiple-access network in Figure~\ref{Figure :  Multiple-access}. In this network, according to Theorem~\ref{Theorem: Multiple-access}, to simulate two pairs of maximally entangled qubits, $\ket{\omega_{ABC_1 C_2}}=\ket{\Phi_{AC_1}}\otimes \ket{\Phi_{BC_2}}$,  the quantum communication rate at each relevant link must
satisfy $Q_{i\to j}\geq 1$, reflecting the coordination requirement from the repeater, Charlie, to transmit one qubit to  Alice and one to Bob. Once Charlie performs a local Bell measurement on his two systems $C_1$ and $C_2$, entanglement is swapped, and Alice and Bob are left with a maximally entangled state, as discussed in Remark~\ref{Remark: repeaters}. 
In this setting, the quantum repeater converts the correlations established by the coordination protocol between Alice and Charlie and between Bob and Charlie into maximal entanglement between Alice and Bob.
This repeater-assisted simulation can generate a broad range of quantum correlations, with the optimal communication rates specified in Theorem~\ref{Theorem: Multiple-access}.

\subsection{Summary and Future Directions}
\label{Subsection: Summary and Future Directions}

The study of quantum coordination enables a deeper understanding of distributed quantum systems, in which the objective extends beyond the reliable transmission of information and instead focuses on how users cooperate to generate desired correlations among remote systems. 
The coordination framework unifies a broad range of information-theoretic tasks that have traditionally been studied independently. Such tasks include quantum channel simulation \cite{berta2011quantum}, channel resolvability and soft covering \cite{atif2024quantum}, state merging and state redistribution \cite{anshu2018one}, entanglement dilution and distillation \cite{chitambar2019quantum}, and distributed quantum source coding \cite{wakakuwa2022one}. 
Within the coordination framework, these protocols can be viewed as methods for generating, manipulating, or distributing quantum correlations under communication constraints.

In this review, we surveyed the foundations of both strong and empirical coordination in quantum networks. 
Strong coordination requires simulating a memoryless quantum source and is closely related to channel simulation, resolvability, soft covering, and reverse Shannon theorems \cite{fang2018quantum,Bloch_resolvability_2019,hayashi2025resolvability}. 
Empirical coordination imposes a weaker requirement, focusing instead on the average state generated by the network and the observed statistics obtained from repeated experiments. 

We further examined coordination over two-node, cascade, broadcast, multiple-access, and no-communication networks. The resulting characterizations show the trade-off between communication requirements and pre-shared resources, illustrating that, in some cases, common randomness and entanglement assistance can reduce the communication rates needed to generate quantum correlations. 
The rate characterizations emphasize the fundamental difference between the quantum and classical settings, as we saw in the cascade network in Examples~\ref{Mixed_example} and \ref{Pure_example}. They also demonstrate that the network topology fundamentally influences the types of correlations that can be produced; see Remarks~\ref{MAC_product_state}, \ref{Remark:Entanglement_BC}, \ref{Remark:Entanglement_2}. 

Beyond its theoretical significance, coordination also provides a useful perspective on several applications. 
In nonlocal games, the available communication and entanglement resources determine the correlations available to the players and therefore directly affect their achievable winning probabilities. 
In device-independent quantum key distribution, security is certified by the violation of a Bell inequality and the generation of nonlocal correlations. 
Another important application arises in quantum repeater networks, where coordination governs the generation and distribution of long-distance entanglement. 
More broadly, coordination is expected to play an increasingly important role in distributed sensing, cooperative quantum systems, future quantum internet architectures, and task-oriented communication networks.

Despite the substantial progress achieved in recent years, many important challenges remain open. 
We conclude by highlighting several promising directions for future research.

\begin{itemize}

\item \emph{General quantum network coordination:}
Develop coordination theorems for arbitrary network topologies comprising classical links, quantum links, common randomness, entanglement assistance, and quantum side information. A central open problem is the characterization of coordination capacity regions for general quantum networks, analogous to the role of network information theory in classical communication systems.
How intermediate nodes should process, distribute, and consume entanglement remains largely unresolved~\cite{pereg2024quantum}.
Another direction is to investigate the benefits of two-way communication, feedback, adaptive protocols, and dynamic network structures. 
Determining the communication and entanglement costs of generating and converting between states remains an important challenge~\cite{salek2023new}.

\item \emph{Coordination over noisy quantum resources:}
Most existing results assume noiseless communication links or ideal entanglement. Future work should address strong and empirical coordination over noisy quantum channels, imperfect entanglement resources, decohering quantum memories, and realistic quantum-network hardware.

\item \emph{One-shot and finite-blocklength coordination:}
Develop non-asymptotic achievability and converse bounds, second-order asymptotics, moderate-deviation analyses, and finite-blocklength trade-off among communication, entanglement, randomness, latency, and approximation accuracy \cite{datta2015secondorder,ramakrishnan2023moderate}.

\item \emph{Nonlocal games and device-independent quantum information processing:}
Establish deeper connections between coordination theory, Bell nonlocality, and nonlocal games. Open questions include the communication, randomness, and entanglement costs required to simulate nonlocal correlations, achieve optimal quantum strategies, and characterize multipartite nonlocal behavior \cite{regev2009simulating,miller2017universal}.
Another promising direction is to apply coordination-theoretic tools to DI-QKD, randomness amplification, and self-testing protocols. A systematic coordination framework for device-independent tasks remains to be developed \cite{vazirani2014fully}.

\item \emph{Distributed quantum sensing and computation:}
Study coordination requirements in distributed sensing, quantum computing, and networked quantum information processing, where multiple users must jointly generate and exploit quantum correlations to perform a common task \cite{matera2016coherent}.

\item \emph{Task-oriented quantum coordination:}
An important direction is to study coordination problems in which distributed quantum users cooperate to perform inference, estimation, optimization, or learning tasks. In such settings, the coordination criterion can be defined in terms of the performance of the underlying task rather than only through the simulation of a desired state \cite{mostaani2022task}. This raises the question of how much communication, common randomness, entanglement, and measurement resources are needed for distributed quantum protocols \cite{xuanqiang2021practical}. Relevant examples include collaborative quantum learning \cite{xia2021quantumfed}, multi-agent quantum decision making \cite{yun2022quantum}, and distributed quantum sensing \cite{proctor2018multiparameter}. Strong and empirical coordination criteria may then be developed according to the requirements of each task.

\end{itemize}

Ultimately, coordination provides a common language for understanding how distributed quantum systems generate, manipulate, and exploit correlations under resource constraints. 
As quantum technologies continue to mature, coordination theory is expected to become an increasingly important component of quantum communication, cryptography, networking, and distributed quantum information processing.

\bibliography{references}
\end{document}

%% file: tikz_pack.tex
\usepackage[utf8]{inputenc}
\usepackage{mathrsfs}

\usepackage{accents}
\newlength{\dhatheight}

\usepackage{lscape}

\usepackage{tikz}
\usetikzlibrary{positioning}
\usetikzlibrary{decorations.markings}
\usetikzlibrary{arrows}
\usetikzlibrary{arrows.meta}
\usepackage[free-standing-units]{siunitx}
\usepackage[siunitx, RPvoltages]{circuitikz}

\makeatletter
\ctikzset{bipoles/my switch/height/.initial=.5}
\ctikzset{bipoles/my switch/width/.initial=.50}
\pgfcircdeclarebipole{}{}{myswitch}{%
\ctikzvalof{bipoles/my switch/height}}{\ctikzvalof{bipoles/my switch/width}}{
        \pgfsetlinewidth{\pgfkeysvalueof{/tikz/circuitikz/bipoles/thickness}\pgfstartlinewidth}
        \pgfpathmoveto{\pgfpoint{\pgf@circ@res@left}{0pt}}
        \pgfpathlineto{\pgfpoint{\pgf@circ@res@right}{.75\pgf@circ@res@up}}
        \pgfusepath{draw}
        \pgfsetlinewidth{\pgfstartlinewidth}        
        \pgftransformshift{\pgfpoint{\pgf@circ@res@left}{0pt}}
        \pgfnode{ocirc}{center}{}{}{\pgfusepath{draw}}
        \pgftransformshift{\pgfpoint{2\pgf@circ@res@right}{0pt}}
        \pgfnode{ocirc}{center}{}{}{\pgfusepath{draw}}
}
\def\pgf@circ@myswitch@path#1{\pgf@circ@bipole@path{myswitch}{#1}}
\compattikzset{my switch/.style = {\circuitikzbasekey, 
/tikz/to path=\pgf@circ@myswitch@path}}

\usepackage{xcolor}
\definecolor{mypurple}{rgb}{1,0,1}
	\definecolor{apricot}{rgb}{0.98, 0.81, 0.69}
		\definecolor{azure}{rgb}{0.0, 0.5, 1.0}
			\definecolor{darkmidnightblue}{rgb}{0.0, 0.2, 0.4}
				\definecolor{tearose}{rgb}{0.97, 0.51, 0.47}
					\definecolor{teagreen}{rgb}{0.82, 0.94, 0.75}
						\definecolor{indigo}{rgb}{0.29, 0.0, 0.51}
							\definecolor{tyrianpurple}{rgb}{0.4, 0.01, 0.24}

\tikzstyle{myedgestyle} = [-triangle 60]
\tikzstyle{block} = [draw, shape=rectangle, minimum height=3cm, minimum width=3cm, node distance=4cm, line width=0.5pt]
\tikzstyle{sum} = [draw, shape=circle, node distance=4cm, line width=0.5pt, minimum width=2.em]
\tikzstyle{mult} = [draw, shape=circle, node distance=3cm, line width=0.5pt, minimum width=2.em]
\tikzstyle{branch}=[fill,shape=circle,minimum size=4cm,inner sep=0pt]

%% file: arXiv.bbl
\ifdefined\bibstar\else\newcommand{\bibstar}[1]{}\fi
\begin{thebibliography}{100}
\providecommand{\url}[1]{#1}
\csname url@samestyle\endcsname
\providecommand{\newblock}{\relax}
\providecommand{\bibinfo}[2]{#2}
\providecommand{\BIBentrySTDinterwordspacing}{\spaceskip=0pt\relax}
\providecommand{\BIBentryALTinterwordstretchfactor}{4}
\providecommand{\BIBentryALTinterwordspacing}{\spaceskip=\fontdimen2\font plus
\BIBentryALTinterwordstretchfactor\fontdimen3\font minus \fontdimen4\font\relax}
\providecommand{\BIBforeignlanguage}[2]{{%
\expandafter\ifx\csname l@#1\endcsname\relax
\typeout{** WARNING: IEEEtran.bst: No hyphenation pattern has been}%
\typeout{** loaded for the language `#1'. Using the pattern for}%
\typeout{** the default language instead.}%
\else
\language=\csname l@#1\endcsname
\fi
#2}}
\providecommand{\BIBdecl}{\relax}
\BIBdecl

\bibitem{cuff2010coordination}
P.~W. Cuff, H.~H. Permuter, and T.~M. Cover, ``Coordination capacity,'' \emph{IEEE Trans. Inf. Theory}, vol.~56, no.~9, pp. 4181--4206, 2010.

\bibitem{sudan2019communication}
M.~Sudan, H.~Tyagi, and S.~Watanabe, ``Communication for generating correlation: A unifying survey,'' \emph{IEEE Trans. Inf. Theory}, vol.~66, no.~1, pp. 5--37, 2019.

\bibitem{stankovic2003real}
J.~A. Stankovic, T.~Abdelzaher, C.~Lu, L.~Sha, and J.~C. Hou, ``Real-time communication and coordination in embedded sensor networks,'' \emph{Proc. IEEE}, vol.~91, no.~7, pp. 1002--1022, 2003.

\bibitem{ahangar2021survey}
M.~N. Ahangar, Q.~Z. Ahmed, F.~A. Khan, and M.~Hafeez, ``A survey of autonomous vehicles: Enabling communication technologies and challenges,'' \emph{Sensors}, vol.~21, no.~3, p. 706, 2021.

\bibitem{cuff2011coordination}
P.~Cuff and L.~Zhao, ``Coordination using implicit communication,'' in \emph{2011 IEEE Inf. Theory Workshop (ITW 2011)}.\hskip 1em plus 0.5em minus 0.4em\relax IEEE, 2011, pp. 467--471.

\bibitem{borcea2002cooperative}
C.~Borcea, D.~Iyer, P.~Kang, A.~Saxena, and L.~Iftode, ``Cooperative computing for distributed embedded systems,'' in \emph{Proc. 22nd Int. Conf. Distrib. Comput. Syst.}\hskip 1em plus 0.5em minus 0.4em\relax IEEE, 2002, pp. 227--236.

\bibitem{mylonakis2020remote}
M.~Mylonakis, P.~A. Stavrou, and M.~Skoglund, ``Remote empirical coordination,'' in \emph{2020 Int. Symp. Inf. Theory Appl. (ISITA 2020)}.\hskip 1em plus 0.5em minus 0.4em\relax IEEE, 2020, pp. 31--35.

\bibitem{he2020internet}
L.~He, M.~Xue, and B.~Gu, ``{I}nternet-of-{T}hings enabled supply chain planning and coordination with big data services: Certain theoretic implications,'' \emph{J. Manage. Sci. Eng.}, vol.~5, no.~1, pp. 1--22, 2020.

\bibitem{torres2023message}
L.~Torres-Figueroa, R.~Ferrara, C.~Deppe, and H.~Boche, ``Message identification for task-oriented communications: Exploiting an exponential increase in the number of connected devices,'' \emph{IEEE Internet of Things Mag.}, vol.~6, no.~4, pp. 42--47, 2023.

\bibitem{10619085}
M.~Lederman and U.~Pereg, ``Secure communication with unreliable entanglement assistance,'' in \emph{2024 IEEE Int. Symp. Inf. Theory (ISIT)}, 2024, pp. 1017--1022.

\bibitem{ElGamalKim:11b}
A.~E. Gamal and Y.-H. Kim, \emph{Network Information Theory}.\hskip 1em plus 0.5em minus 0.4em\relax Cambridge University Press, 2011.

\bibitem{tse1999linear}
D.~N.~C. Tse and S.~V. Hanly, ``Linear multiuser receivers: Effective interference, effective bandwidth and user capacity,'' \emph{IEEE Trans. Inf. theory}, vol.~45, no.~2, pp. 641--657, 1999.

\bibitem{rosenberger2023identification}
J.~Rosenberger, C.~Deppe, and U.~Pereg, ``Identification over quantum broadcast channels,'' \emph{Quantum Inf. Process.}, vol.~22, no.~10, p. 361, 2023.

\bibitem{pereg2023multiple}
U.~Pereg, C.~Deppe, and H.~Boche, ``The multiple-access channel with entangled transmitters,'' in \emph{Proc. Global Commun. Conf. (GLOBECOM'2023)}.\hskip 1em plus 0.5em minus 0.4em\relax IEEE, 2023, pp. 3173--3178.

\bibitem{8768393}
K.~S.~K. Arumugam and M.~R. Bloch, ``Covert communication over a $k$ -user multiple-access channel,'' \emph{IEEE Trans. Inf. Theory}, vol.~65, no.~11, pp. 7020--7044, 2019.

\bibitem{le2017joint}
M.~Le~Treust, ``Joint empirical coordination of source and channel,'' \emph{IEEE Trans. Inf. Theory}, vol.~63, no.~8, pp. 5087--5114, 2017.

\bibitem{notzel2020entanglement}
J.~N{\"o}tzel, ``Entanglement-enabled communication,'' \emph{IEEE J. Sel. Areas Inf. Theory}, vol.~1, no.~2, pp. 401--415, 2020.

\bibitem{9232550}
J.~N\"otzel and S.~DiAdamo, ``Entanglement-enabled communication for the {I}nternet of things,'' in \emph{2020 Int. Conf. Comput., Inf. Telecommun. Syst. (CITS)}, 2020, pp. 1--6.

\bibitem{burenkov2021practical}
I.~Burenkov, M.~Jabir, and S.~Polyakov, ``Practical quantum-enhanced receivers for classical communication,'' \emph{AVS quantum sci.}, vol.~3, no.~2, 2021.

\bibitem{granelli2022novel}
F.~Granelli, R.~Bassoli, J.~N{\"o}tzel, F.~H. Fitzek, H.~Boche, N.~L. da~Fonseca \emph{et~al.}, ``A novel architecture for future classical-quantum communication networks,'' \emph{Wireless Commun. Mobile Comput.}, vol. 2022, 2022.

\bibitem{natur2025quantum}
H.~Natur and U.~Pereg, ``Quantum coordination rates in multi-user networks,'' \emph{IEEE Trans. Inf. Theory}, 2025.

\bibitem{HsiehWilde:10p1}
M.~{Hsieh} and M.~M. {Wilde}, ``Trading classical communication, quantum communication, and entanglement in quantum {S}hannon theory,'' \emph{IEEE Trans. Inf. Theory}, vol.~56, no.~9, pp. 4705--4730, Sep 2010.

\bibitem{shapiro2020quantum}
J.~H. Shapiro, ``The quantum illumination story,'' \emph{IEEE Aerospace and Electronic Systems Magazine}, vol.~35, no.~4, pp. 8--20, 2020.

\bibitem{myerson1997game}
R.~B. Myerson, \emph{Game Theory: Analysis of Conflict}.\hskip 1em plus 0.5em minus 0.4em\relax Cambridge, MA: Harvard University Press, 1997.

\bibitem{9706458}
U.~Pereg, C.~Deppe, and H.~Boche, ``The quantum multiple-access channel with cribbing encoders,'' \emph{IEEE Trans. Inf. Theory}, vol.~68, no.~6, pp. 3965--3988, 2022.

\bibitem{brunner2014bell}
N.~Brunner, D.~Cavalcanti, S.~Pironio, V.~Scarani, and S.~Wehner, ``{B}ell nonlocality,'' \emph{Rev. modern phys.}, vol.~86, no.~2, pp. 419--478, 2014.

\bibitem{buhrman2010nonlocality}
H.~Buhrman, R.~Cleve, S.~Massar, and R.~de~Wolf, ``Nonlocality and communication complexity,'' \emph{Reviews of Modern Physics}, vol.~82, no.~1, pp. 665--698, 2010.

\bibitem{clauser1969proposed}
J.~F. Clauser, M.~A. Horne, A.~Shimony, and R.~A. Holt, ``Proposed experiment to test local hidden-variable theories,'' \emph{Phys. rev. lett.}, vol.~23, no.~15, p. 880, 1969.

\bibitem{bell1964einstein}
J.~S. Bell, ``On the {E}instein {P}odolsky {R}osen paradox,'' \emph{Phys. Phys. Fiz.}, vol.~1, no.~3, p. 195, 1964.

\bibitem{blum1983coin}
M.~Blum, ``Coin flipping by telephone a protocol for solving impossible problems,'' \emph{ACM SIGACT News}, vol.~15, no.~1, pp. 23--27, 1983.

\bibitem{crepeau2020commitment}
C.~Cr{\'e}peau, R.~Dowsley, and A.~C. Nascimento, ``On the commitment capacity of unfair noisy channels,'' \emph{IEEE Transactions on Information Theory}, vol.~66, no.~6, pp. 3745--3752, 2020.

\bibitem{goldreich2019play}
O.~Goldreich, S.~Micali, and A.~Wigderson, ``How to play any mental game, or a completeness theorem for protocols with honest majority,'' in \emph{Providing Sound Foundations for Cryptography: On the Work of Shafi Goldwasser and Silvio Micali}, 2019, pp. 307--328.

\bibitem{chaum1988multiparty}
D.~Chaum, I.~B. Damg{\aa}rd, and J.~Van~de Graaf, ``Multiparty computations ensuring privacy of each party’s input and correctness of the result,'' in \emph{Advances in Cryptology—CRYPTO’87: Proceedings 7}.\hskip 1em plus 0.5em minus 0.4em\relax Springer, 1988, pp. 87--119.

\bibitem{even1985randomized}
S.~Even, O.~Goldreich, and A.~Lempel, ``A randomized protocol for signing contracts,'' \emph{Communications of the ACM}, vol.~28, no.~6, pp. 637--647, 1985.

\bibitem{goldreich1991proofs}
O.~Goldreich, S.~Micali, and A.~Wigderson, ``Proofs that yield nothing but their validity or all languages in np have zero-knowledge proof systems,'' \emph{Journal of the ACM (JACM)}, vol.~38, no.~3, pp. 690--728, 1991.

\bibitem{brassard1988minimum}
G.~Brassard, D.~Chaum, and C.~Cr{\'e}peau, ``Minimum disclosure proofs of knowledge,'' \emph{Journal of computer and system sciences}, vol.~37, no.~2, pp. 156--189, 1988.

\bibitem{lo1997quantum}
H.-K. Lo and H.~F. Chau, ``Is quantum bit commitment really possible?'' \emph{Phys. Rev. Lett.}, vol.~78, no.~17, p. 3410, 1997.

\bibitem{mayers1997unconditionally}
D.~Mayers, ``Unconditionally secure quantum bit commitment is impossible,'' \emph{Phys. rev. lett.}, vol.~78, no.~17, p. 3414, 1997.

\bibitem{lo1998quantum}
H.-K. Lo and H.~F. Chau, ``Why quantum bit commitment and ideal quantum coin tossing are impossible,'' \emph{Physica D: Nonlinear Phenomena}, vol. 120, no. 1-2, pp. 177--187, 1998.

\bibitem{winter2003commitment}
A.~Winter, A.~C. Nascimento, and H.~Imai, ``Commitment capacity of discrete memoryless channels,'' in \emph{Cryptography and Coding: 9th IMA Int. Conf., Cirencester, UK, December 16-18, 2003. Proc. 9}.\hskip 1em plus 0.5em minus 0.4em\relax Springer, 2003, pp. 35--51.

\bibitem{bennett1984quantum}
C.~H. Bennett and G.~Brassard, ``Quantum cryptography: Public key distribution and coin tossing,'' in \emph{Proceedings of the IEEE International Conference on Computers, Systems and Signal Processing}, Bangalore, India, 1984, pp. 175--179.

\bibitem{ekert1991quantum}
A.~K. Ekert, ``Quantum cryptography based on bell’s theorem,'' \emph{Physical review letters}, vol.~67, no.~6, p. 661, 1991.

\bibitem{acin2007device}
A.~Ac{\'\i}n, N.~Brunner, N.~Gisin, S.~Massar, S.~Pironio, and V.~Scarani, ``Device-independent security of quantum cryptography against collective attacks,'' \emph{Phys. Rev. Lett.}, vol.~98, no.~23, p. 230501, 2007.

\bibitem{pironio2009device}
S.~Pironio, A.~Ac{\'\i}n, N.~Brunner, N.~Gisin, S.~Massar, and V.~Scarani, ``Device-independent quantum key distribution secure against collective attacks,'' \emph{New J. Phys.}, vol.~11, no.~4, p. 045021, 2009.

\bibitem{devetak2005distillation}
I.~Devetak and A.~Winter, ``Distillation of secret key and entanglement from quantum states,'' \emph{Proc. Royal Society A: Math. Phys. Engin. Scien.}, vol. 461, no. 2053, pp. 207--235, 2005.

\bibitem{wolf2021quantum}
R.~Wolf, ``Quantum key distribution protocols,'' in \emph{Quantum Key Distribution: An Introduction with Exercises}.\hskip 1em plus 0.5em minus 0.4em\relax Springer, 2021, pp. 91--116.

\bibitem{wyner1975common}
A.~Wyner, ``The common information of two dependent random variables,'' \emph{IEEE Transactions on Information Theory}, vol.~21, no.~2, pp. 163--179, 1975.

\bibitem{cuff2008communication}
P.~Cuff, ``Communication requirements for generating correlated random variables,'' in \emph{2008 IEEE Int. Symp. Inf. Theory}, 2008, pp. 1393--1397.

\bibitem{fuchs1996quantum}
C.~A. Fuchs and A.~Peres, ``Quantum-state disturbance versus information gain: Uncertainty relations for quantum information,'' \emph{Phys. Rev. A}, vol.~53, no.~4, p. 2038, 1996.

\bibitem{fuchs2000quantum}
------, ``Quantum theory needs no ‘interpretation’,'' \emph{Phys. today}, vol.~53, no.~3, pp. 70--71, 2000.

\bibitem{bricmont2016making}
J.~Bricmont, \emph{Making sense of quantum mechanics}.\hskip 1em plus 0.5em minus 0.4em\relax Springer, 2016, vol.~37.

\bibitem{bennett2002entanglement}
C.~H. Bennett, P.~W. Shor, J.~A. Smolin, and A.~V. Thapliyal, ``Entanglement-assisted capacity of a quantum channel and the reverse {S}hannon theorem,'' \emph{IEEE trans. Inf. Theory}, vol.~48, no.~10, pp. 2637--2655, 2002.

\bibitem{kumagai2016second}
W.~Kumagai and M.~Hayashi, ``Second-order asymptotics of conversions of distributions and entangled states based on rayleigh-normal probability distributions,'' \emph{IEEE Trans. Inf. Theory}, vol.~63, no.~3, pp. 1829--1857, 2016.

\bibitem{schumacher1995quantum}
B.~Schumacher, ``Quantum coding,'' \emph{Phys. Rev. A}, vol.~51, no.~4, p. 2738, 1995.

\bibitem{natur2025empirical_ITW}
H.~Natur and U.~Pereg, ``Empirical coordination of quantum correlations,'' in \emph{2025 IEEE Information Theory Workshop (ITW)}.\hskip 1em plus 0.5em minus 0.4em\relax IEEE, 2025, pp. 1--7.

\bibitem{bennett2014quantum}
C.~H. Bennett, I.~Devetak, A.~W. Harrow, P.~W. Shor, and A.~Winter, ``The quantum reverse {S}hannon theorem and resource tradeoffs for simulating quantum channels,'' \emph{IEEE Trans. Inf. Theory}, vol.~60, no.~5, pp. 2926--2959, 2014.

\bibitem{hayashi2016quantum}
M.~Hayashi, \emph{Quantum Information Theory: Mathematical Foundation}.\hskip 1em plus 0.5em minus 0.4em\relax Springer, 2016.

\bibitem{hayden2003communication}
P.~Hayden and A.~Winter, ``Communication cost of entanglement transformations,'' \emph{Phys. Rev. A}, vol.~67, no.~1, p. 012326, 2003.

\bibitem{harrow2004tight}
A.~W. Harrow and H.-K. Lo, ``A tight lower bound on the classical communication cost of entanglement dilution,'' \emph{IEEE Trans. Inf. Theory}, vol.~50, no.~2, pp. 319--327, 2004.

\bibitem{horodecki2005partial}
M.~Horodecki, J.~Oppenheim, and A.~Winter, ``Partial quantum information,'' \emph{Nature}, vol. 436, no. 7051, pp. 673--676, 2005.

\bibitem{horodecki2007quantum}
------, ``Quantum state merging and negative information,'' \emph{Commun. Math. Phys.}, vol. 269, pp. 107--136, 2007.

\bibitem{devetak2008exact}
I.~Devetak and J.~Yard, ``Exact cost of redistributing multipartite quantum states,'' \emph{Phys. Rev. Lett.}, vol. 100, no.~23, p. 230501, 2008.

\bibitem{Yard_Devetak_2009}
J.~T. Yard and I.~Devetak, ``Optimal quantum source coding with quantum side information at the encoder and decoder,'' \emph{IEEE Trans. Inf. Theory}, vol.~55, no.~11, pp. 5339--5351, 2009.

\bibitem{luo2009channel}
Z.~Luo and I.~Devetak, ``Channel simulation with quantum side information,'' \emph{IEEE Trans. Inf. Theory}, vol.~55, no.~3, pp. 1331--1342, 2009.

\bibitem{abeyesinghe2009mother}
A.~Abeyesinghe, I.~Devetak, P.~Hayden, and A.~Winter, ``The mother of all protocols: Restructuring quantum information’s family tree,'' \emph{Proc. Roy. Soc. A: Math., Phys. Eng. Sci.}, vol. 465, no. 2108, pp. 2537--2563, 2009.

\bibitem{ahn2006distributed}
C.~Ahn, A.~C. Doherty, P.~Hayden, and A.~J. Winter, ``On the distributed compression of quantum information,'' \emph{IEEE transactions on information theory}, vol.~52, no.~10, pp. 4349--4357, 2006.

\bibitem{9039682}
Z.~Baghali~Khanian and A.~Winter, ``Distributed compression of correlated classical-quantum sources or: The price of ignorance,'' \emph{IEEE Trans. Inf. Theory}, vol.~66, no.~9, pp. 5620--5633, 2020.

\bibitem{salek2018quantum}
S.~Salek, D.~Cadamuro, P.~Kammerlander, and K.~Wiesner, ``Quantum rate-distortion coding of relevant information,'' \emph{IEEE Trans. Inf. Theory}, vol.~65, no.~4, pp. 2603--2613, 2018.

\bibitem{Faithful_simulation_Heidari_2019}
T.~A. Atif, M.~Heidari, and S.~S. Pradhan, ``Faithful simulation of distributed quantum measurements with applications in distributed rate-distortion theory,'' \emph{IEEE Trans. Inf. Theory}, vol.~68, no.~2, pp. 1085--1118, 2022.

\bibitem{streltsov2020rates}
A.~Streltsov, C.~Meignant, and J.~Eisert, ``Rates of multipartite entanglement transformations,'' \emph{Phys. Rev. Lett.}, vol. 125, no.~8, p. 080502, 2020.

\bibitem{george2024coherent}
I.~George and H.-C. Cheng, ``Coherent distributed source simulation as multipartite quantum state splitting,'' in \emph{2024 IEEE Int. Symp. Inf. Theory (ISIT)}.\hskip 1em plus 0.5em minus 0.4em\relax IEEE, 2024, pp. 1221--1226.

\bibitem{cheng2023quantum}
H.-C. Cheng, L.~Gao, and M.~Berta, ``Quantum broadcast channel simulation via multipartite convex splitting,'' \emph{arXiv preprint arXiv:2304.12056}, 2023.

\bibitem{cao2024channel}
M.~X. Cao, N.~Ramakrishnan, M.~Berta, and M.~Tomamichel, ``Channel simulation: Finite blocklengths and broadcast channels,'' \emph{IEEE Trans. Inf. Theory}, 2024.

\bibitem{nema2024one}
A.~Nema, S.~Sreekumar, and M.~Berta, ``One-shot multiple access channel simulation,'' in \emph{2024 IEEE Int. Symp. Inf. Theory (ISIT)}.\hskip 1em plus 0.5em minus 0.4em\relax IEEE, 2024, pp. 2981--2986.

\bibitem{cuff2013distributed}
P.~Cuff, ``Distributed channel synthesis,'' \emph{IEEE Transactions on Information Theory}, vol.~59, no.~11, pp. 7071--7096, 2013.

\bibitem{wilde2017quantum}
M.~M. Wilde, \emph{Quantum Information Theory}, 2nd~ed.\hskip 1em plus 0.5em minus 0.4em\relax Cambridge University Press, 2017.

\bibitem{einstein1935can}
A.~Einstein, B.~Podolsky, and N.~Rosen, ``Can quantum-mechanical description of physical reality be considered complete?'' \emph{Phys. rev.}, vol.~47, no.~10, p. 777, 1935.

\bibitem{bell1966problem}
J.~S. Bell, ``On the problem of hidden variables in quantum mechanics,'' \emph{Rev. Modern phys.}, vol.~38, no.~3, p. 447, 1966.

\bibitem{slofstra2019set}
W.~Slofstra, ``The set of quantum correlations is not closed,'' in \emph{Forum Math., Pi}, vol.~7.\hskip 1em plus 0.5em minus 0.4em\relax Cambridge University Press, 2019, p.~e1.

\bibitem{cirel1980quantum}
B.~S. Cirel'son, ``Quantum generalizations of {B}ell's inequality,'' \emph{Lett. Math. Phys.}, vol.~4, pp. 93--100, 1980.

\bibitem{brassard2005quantum}
G.~Brassard, A.~Broadbent, and A.~Tapp, ``Quantum pseudo-telepathy,'' \emph{Found. Phys.}, vol.~35, pp. 1877--1907, 2005.

\bibitem{slofstra2018entanglement}
W.~Slofstra and T.~Vidick, ``Entanglement in non-local games and the hyperlinear profile of groups,'' \emph{Annales Henri Poincar{\'e}}, vol.~19, no.~10, pp. 2979--3005, 2018.

\bibitem{dilley2018more}
D.~Dilley and E.~Chitambar, ``More nonlocality with less entanglement in clauser-horne-shimony-holt experiments using inefficient detectors,'' \emph{Physical Review A}, vol.~97, no.~6, p. 062313, 2018.

\bibitem{devetak2003classical}
I.~Devetak and A.~Winter, ``Classical data compression with quantum side information,'' \emph{Phys. Rev. A}, vol.~68, no.~4, p. 042301, 2003.

\bibitem{shannon1948mathematical}
C.~E. Shannon, ``A mathematical theory of communication,'' \emph{The Bell system technical journal}, vol.~27, no.~3, pp. 379--423, 1948.

\bibitem{natur2025empirical}
H.~Natur and U.~Pereg, ``Empirical coordination of separable quantum correlations,'' \emph{AIMS Mathematics}, vol.~10, no.~4, 2025.

\bibitem{han1993approximation}
T.~S. Han and S.~Verd{\'u}, ``Approximation theory of output statistics,'' \emph{IEEE Transactions on Information Theory}, vol.~39, no.~3, pp. 752--772, 1993.

\bibitem{hayashi2006general}
M.~Hayashi, ``General formulas for fixed-length quantum entanglement concentration,'' \emph{IEEE Transactions on Information Theory}, vol.~52, no.~5, pp. 1904--1921, 2006.

\bibitem{atif2023quantum}
T.~A. Atif, S.~S. Pradhan, and A.~Winter, ``Quantum soft-covering lemma with applications to rate-distortion coding, resolvability and identification via quantum channels,'' \emph{arXiv preprint arXiv:2306.12416}, 2023.

\bibitem{berta2016smooth}
M.~Berta, M.~Christandl, and D.~Touchette, ``Smooth entropy bounds on one-shot quantum state redistribution,'' \emph{IEEE Transactions on Information Theory}, vol.~62, no.~3, pp. 1425--1439, 2016.

\bibitem{slepian1973noiseless}
D.~Slepian and J.~Wolf, ``Noiseless coding of correlated information sources,'' \emph{IEEE Trans. inf. Theory}, vol.~19, no.~4, pp. 471--480, 1973.

\bibitem{wyner1976rate}
A.~Wyner and J.~Ziv, ``The rate-distortion function for source coding with side information at the decoder,'' \emph{IEEE Trans. inf. Theory}, vol.~22, no.~1, pp. 1--10, 1976.

\bibitem{yassaee2014achievability}
M.~H. Yassaee, M.~R. Aref, and A.~Gohari, ``Achievability proof via output statistics of random binning,'' \emph{IEEE Trans. Inf. Theory}, vol.~60, no.~11, pp. 6760--6786, 2014.

\bibitem{cervia2020strong}
G.~Cervia, L.~Luzzi, M.~Le~Treust, and M.~R. Bloch, ``Strong coordination of signals and actions over noisy channels with two-sided state information,'' \emph{IEEE Trans. Inf. Theory}, vol.~66, no.~8, pp. 4681--4708, 2020.

\bibitem{jozsa1994new}
R.~Jozsa and B.~Schumacher, ``A new proof of the quantum noiseless coding theorem,'' \emph{J. Modern Opt.}, vol.~41, no.~12, pp. 2343--2349, 1994.

\bibitem{barnum1996general}
H.~Barnum, C.~A. Fuchs, R.~Jozsa, and B.~Schumacher, ``General fidelity limit for quantum channels,'' \emph{Phys. Rev. A}, vol.~54, no.~6, p. 4707, 1996.

\bibitem{GourWildeBrandsenGeng:22a}
G.~Gour, M.~M. Wilde, S.~Brandsen, and I.~J. Geng, ``Inevitability of knowing less than nothing,'' \emph{\textup{\texttt{2208.14424}}}, 2022.

\bibitem{george2023one}
I.~George, M.-H. Hsieh, and E.~Chitambar, ``One-shot bounds on state generation using correlated resources and local encoders,'' in \emph{2023 IEEE Int. Symp. Inf. Theory (ISIT)}.\hskip 1em plus 0.5em minus 0.4em\relax IEEE, 2023, pp. 96--101.

\bibitem{munro2015inside}
W.~J. Munro, K.~Azuma, K.~Tamaki, and K.~Nemoto, ``Inside quantum repeaters,'' \emph{IEEE Journal of Selected Topics in Quantum Electronics}, vol.~21, no.~3, pp. 78--90, 2015.

\bibitem{pereg2023communication}
U.~Pereg, C.~Deppe, and H.~Boche, ``Communication with unreliable entanglement assistance,'' \emph{IEEE Trans. Inf. Theory}, vol.~69, no.~7, pp. 4579--4599, 2023.

\bibitem{wootters1982single}
W.~K. Wootters and W.~H. Zurek, ``A single quantum cannot be cloned,'' \emph{Nature}, vol. 299, no. 5886, pp. 802--803, 1982.

\bibitem{bennett1993teleporting}
C.~H. Bennett, G.~Brassard, C.~Cr{\'e}peau, R.~Jozsa, A.~Peres, and W.~K. Wootters, ``Teleporting an unknown quantum state via dual classical and einstein-podolsky-rosen channels,'' \emph{Physical Review Letters}, vol.~70, no.~13, pp. 1895--1899, 1993.

\bibitem{briegel1998quantum}
H.-J. Briegel, W.~D{\"u}r, J.~I. Cirac, and P.~Zoller, ``Quantum repeaters: the role of imperfect local operations in quantum communication,'' \emph{Phys. Rev. Lett.}, vol.~81, no.~26, p. 5932, 1998.

\bibitem{sangouard2011quantum}
N.~Sangouard, C.~Simon, H.~de~Riedmatten, and N.~Gisin, ``Quantum repeaters based on atomic ensembles and linear optics,'' \emph{Reviews of Modern Physics}, vol.~83, no.~1, pp. 33--80, 2011.

\bibitem{goodenough2021optimizing}
K.~Goodenough, D.~Elkouss, and S.~Wehner, ``Optimizing repeater schemes for the quantum internet,'' \emph{Phys. Rev. A}, vol. 103, no.~3, p. 032610, 2021.

\bibitem{wilksen2024gate}
S.~Wilksen, F.~Lohof, I.~Willmann, F.~Bopp, M.~Lienhart, C.~Thalacker, J.~Finley, M.~Florian, and C.~Gies, ``Gate-based protocol simulations for quantum repeaters using quantum-dot molecules in switchable electric fields,'' \emph{Adv. Quantum Technol.}, vol.~7, no.~3, p. 2300280, 2024.

\bibitem{neuwirth2021quantum}
J.~Neuwirth, F.~B. Basset, M.~B. Rota, E.~Roccia, C.~Schimpf, K.~D. J{\"o}ns, A.~Rastelli, and R.~Trotta, ``Quantum dot technology for quantum repeaters: from entangled photon generation toward the integration with quantum memories,'' \emph{Materials Quantum Technol.}, vol.~1, no.~4, p. 043001, 2021.

\bibitem{bergerhoff2024quantum}
M.~Bergerhoff, O.~Elshehy, S.~Kucera, M.~Kreis, and J.~Eschner, ``Quantum repeater node with free-space coupled trapped ions,'' \emph{Phys. Rev. A}, vol. 110, no.~3, p. 032603, 2024.

\bibitem{dhara2022multiplexed}
P.~Dhara, N.~M. Linke, E.~Waks, S.~Guha, and K.~P. Seshadreesan, ``Multiplexed quantum repeaters based on dual-species trapped-ion systems,'' \emph{Phys. Rev. A}, vol. 105, no.~2, p. 022623, 2022.

\bibitem{mizuochi2024quantum}
N.~Mizuochi and N.~Morioka, ``Quantum light sources based on color centers in diamond and silicon carbide,'' in \emph{Quantum Photon.}\hskip 1em plus 0.5em minus 0.4em\relax Elsevier, 2024, pp. 339--368.

\bibitem{azuma2023quantum}
K.~Azuma, S.~E. Economou, D.~Elkouss, P.~Hilaire, L.~Jiang, H.-K. Lo, and I.~Tzitrin, ``Quantum repeaters: From quantum networks to the quantum internet,'' \emph{Rev. Modern Phys.}, vol.~95, no.~4, p. 045006, 2023.

\bibitem{rodgers2021materials}
L.~V. Rodgers, L.~B. Hughes, M.~Xie, P.~C. Maurer, S.~Kolkowitz, A.~C. Bleszynski~Jayich, and N.~P. de~Leon, ``Materials challenges for quantum technologies based on color centers in diamond,'' \emph{MRS Bulletin}, vol.~46, no.~7, pp. 623--633, 2021.

\bibitem{van2020extending}
P.~van Loock, W.~Alt, C.~Becher, O.~Benson, H.~Boche, C.~Deppe, J.~Eschner, S.~H{\"o}fling, D.~Meschede, P.~Michler \emph{et~al.}, ``Extending quantum links: modules for fiber-and memory-based quantum repeaters,'' \emph{Adv. quantum technol.}, vol.~3, no.~11, p. 1900141, 2020.

\bibitem{berta2011quantum}
M.~Berta, M.~Christandl, and R.~Renner, ``The quantum reverse {S}hannon theorem based on one-shot information theory,'' \emph{Comm. Math. Phys.}, vol. 306, pp. 579--615, 2011.

\bibitem{atif2024quantum}
T.~A. Atif, S.~S. Pradhan, and A.~Winter, ``Quantum soft-covering lemma with applications to rate-distortion coding, resolvability and identification via quantum channels,'' \emph{International Journal of Quantum Information}, vol.~22, no.~5, p. 2440013, 2024.

\bibitem{anshu2018one}
A.~Anshu, R.~Jain, and N.~A. Warsi, ``A one-shot achievability result for quantum state redistribution,'' \emph{IEEE Transactions on Information Theory}, vol.~64, no.~3, pp. 1425--1435, 2018.

\bibitem{chitambar2019quantum}
E.~Chitambar and G.~Gour, ``Quantum resource theories,'' \emph{Reviews of Modern Physics}, vol.~91, no.~2, p. 025001, 2019.

\bibitem{wakakuwa2022one}
E.~Wakakuwa, Y.~Nakata, and M.-H. Hsieh, ``One-shot hybrid state redistribution,'' \emph{Quantum}, vol.~6, p. 724, 2022.

\bibitem{fang2018quantum}
K.~Fang, X.~Wang, M.~Tomamichel, and M.~Berta, ``Quantum channel simulation and the channel's smooth max-information,'' in \emph{2018 IEEE International Symposium on Information Theory (ISIT)}.\hskip 1em plus 0.5em minus 0.4em\relax IEEE, 2018, pp. 2326--2330.

\bibitem{Bloch_resolvability_2019}
M.~Bloch and J.~N. Laneman, ``Channel resolvability and the soft covering lemma: A comprehensive survey,'' \emph{Foundations and Trends in Communications and Information Theory}, vol.~12, no. 2-3, pp. 219--313, 2015.

\bibitem{hayashi2025resolvability}
M.~Hayashi, H.-C. Cheng, and L.~Gao, ``Resolvability of classical-quantum channels,'' \emph{IEEE Transactions on Information Theory}, vol.~71, no.~8, pp. 6061--6074, 2025.

\bibitem{pereg2024quantum}
U.~Pereg, ``Quantum relay channels,'' \emph{arXiv preprint arXiv:2411.16263}, 2024.

\bibitem{salek2023new}
F.~Salek and A.~Winter, ``New protocols for conference key and multipartite entanglement distillation,'' \emph{arXiv preprint arXiv:2308.01134}, 2023.

\bibitem{datta2015secondorder}
N.~Datta and F.~Leditzky, ``Second-order asymptotics for source coding, dense coding, and pure-state entanglement conversions,'' \emph{IEEE Transactions on Information Theory}, vol.~61, no.~1, pp. 582--608, 2015.

\bibitem{ramakrishnan2023moderate}
N.~Ramakrishnan, M.~Tomamichel, and M.~Berta, ``Moderate deviation expansion for fully quantum tasks,'' \emph{IEEE Transactions on Information Theory}, vol.~69, no.~8, pp. 5041--5059, 2023.

\bibitem{regev2009simulating}
O.~Regev and B.~Toner, ``Simulating quantum correlations with finite communication,'' \emph{SIAM Journal on Computing}, vol.~39, no.~4, pp. 1562--1580, 2009.

\bibitem{miller2017universal}
C.~A. Miller and Y.~Shi, ``Universal security for randomness expansion from the spot-checking protocol,'' \emph{SIAM Journal on Computing}, vol.~46, no.~4, pp. 1304--1335, 2017.

\bibitem{vazirani2014fully}
U.~Vazirani and T.~Vidick, ``Fully device-independent quantum key distribution,'' \emph{Physical Review Letters}, vol. 113, no.~14, p. 140501, 2014.

\bibitem{matera2016coherent}
J.~M. Matera, D.~Egloff, N.~Killoran, and M.~B. Plenio, ``Coherent control of quantum systems as a resource theory,'' \emph{Quantum Sci. Tech.}, vol.~1, no.~1, p. 01LT01, 2016.

\bibitem{mostaani2022task}
A.~Mostaani, T.~X. Vu, S.~K. Sharma, V.-D. Nguyen, Q.~Liao, and S.~Chatzinotas, ``Task-oriented communication design in cyber-physical systems: A survey on theory and applications,'' \emph{IEEE Access}, vol.~10, pp. 133\,843--133\,868, 2022.

\bibitem{xuanqiang2021practical}
Z.~Xuanqiang, Z.~Benchi, Z.~Wang, Z.~Song, and X.~Wang, ``Practical distributed quantum information processing with loccnet,'' \emph{npj Quantum Information}, vol.~7, no.~1, 2021.

\bibitem{xia2021quantumfed}
Q.~Xia and Q.~Li, ``{QuantumFed}: A federated learning framework for collaborative quantum training,'' in \emph{2021 IEEE Global Communications Conference (GLOBECOM)}, 2021, pp. 1--6.

\bibitem{yun2022quantum}
W.~J. Yun, Y.~Kwak, J.~P. Kim, H.~Cho, S.~Jung, J.~Park, and J.~Kim, ``Quantum multi-agent reinforcement learning via variational quantum circuit design,'' in \emph{2022 IEEE 42nd International Conference on Distributed Computing Systems (ICDCS)}, 2022, pp. 1332--1335.

\bibitem{proctor2018multiparameter}
T.~J. Proctor, P.~A. Knott, and J.~A. Dunningham, ``Multiparameter estimation in networked quantum sensors,'' \emph{Phys. Rev. Lett.}, vol. 120, no.~8, p. 080501, 2018.

\end{thebibliography}
